# Buckyball-metal complexes as promising carriers of astronomical unidentified infrared emission bands

Gao-Lei Hou[1,5]*, Olga V. Lushchikova[2], Joost M. Bakker[2], Peter Lievens[1], Leen Decin[3,4]*, Ewald Janssens[1]

[1]Quantum Solid-State Physics, KU Leuven, Celestijnenlaan 200D, 3001 Leuven, Belgium

[2]Radboud University, Institute for Molecules and Materials, FELIX Laboratory, Toernooiveld 7, 6525 ED Nijmegen, The Netherlands

[3]Institute of Astronomy, KU Leuven, Celestijnenlaan 200D, 3001 Leuven, Belgium

[4]School of Chemistry, University of Leeds, Leeds, LS2 9JT, United Kingdom

[5]MOE Key Laboratory for Non-Equilibrium Synthesis and Modulation of Condensed Matter, School of Physics, Xi´an Jiaotong University, Xi´an, 710049 Shaanxi, China

E-mails: gaolei.hou@xjtu.edu.cn (G.-L.H.) or leen.decin@kuleuven.be (L.D.)

**Infrared emission bands with wavelengths between 3–20 μm are observed in a variety of astrophysical environments[1,2]. They were discovered in the 1970s and are generally attributed to organic compounds[3,4]. However, over 40 years of research efforts still leave the source of these emission bands largely unidentified[5-7]. Here, we report the first laboratory infrared (6–25 μm) spectra of gas-phase fullerene-metal complexes, [$C_{60}$-Metal]$^+$ (Metal = Fe and V), and show with density functional theory calculations that complexes of $C_{60}$ with cosmically abundant metals, including Li, Na, K, Mg, Ca, Al, V, and Fe, all have similar infrared spectral patterns. Comparison with observational infrared spectra from several fullerene-rich planetary nebulae demonstrates a strong positive linear cross-correlation. The infrared features of [$C_{60}$-Metal]$^+$ coincide with four bands attributed earlier to neutral $C_{60}$ bands, and in addition also with several to date unexplained bands. Abundance and collision theory estimates furthermore indicate that [$C_{60}$-Metal]$^+$ could plausibly form and survive in astrophysical environments. Hence, [$C_{60}$-Metal]$^+$ are proposed as promising carriers, in supplement to $C_{60}$, of astronomical infrared emission bands, potentially representing the largest molecular species in space other than the bare fullerenes $C_{60}$, $C_{60}^+$, and $C_{70}$. This work opens a new chapter for studying cosmic fullerene species and carbon chemistry in the Universe.**

Astronomical unidentified infrared emission (UIE) bands contain a wealth of information about the physical and chemical conditions in the emitting regions, but their precise carriers are largely unknown, knowledge of which is potentially crucial to probe cosmic star formation history, interstellar chemistry, and galactic evolution[5-7]. Over the past 40 years, general consensus has been reached that UIE features at 3.3, 6.2, 7.7, 8.6, and 11.3 μm can be attributed predominantly to vibrations of polycyclic aromatic hydrocarbon (PAH) molecules, although the exact nature of the emitting PAHs remains elusive[5-11]. Later, an alternative model based on mixed aromatic/aliphatic organic nanoparticles (MAON) from the observation of spectral features at 3.4 and 6.85 μm, characteristic for aliphatic compounds, has been proposed[6]. Still, both the PAH hypothesis and MAON model are under discussion[7,8,12].

In contrast, fullerenes have been confirmed unequivocally as UIE carriers[13-19], notably in the young planetary nebula (PN) Tc 1, in particular through the $C_{60}$ infrared spectral signatures at 7.0, 8.5, 17.4, and 18.9 μm[13]. Its cation, $C_{60}^+$, was speculated to be present in the reflection nebula NGC 7023 from bands at 6.4, 7.1, 8.2, and 10.5 μm[16], before it was firmly identified, through laboratory measurement, as carrier of the diffuse interstellar bands (DIBs) at 9577 and 9632 Å[20-22]. To date, $C_{60}$ and $C_{60}^+$ have been detected in more than 40 cosmic environments[23]. Four planetary nebulae (PNe), Tc 1, SMP LMC 56, SMP SMC 16, and Lin49 are particularly interesting, as their UIE bands show they are $C_{60}$-rich with almost no infrared emissions at wavelengths characteristic for PAHs[13-15].

Although the presence of $C_{60}$ in space is nowadays unambiguous, current knowledge is incomplete as neither fluorescence nor thermal excitation mechanisms can explain the observed band intensity ratios of the four $C_{60}$ features, while the carriers of many other UIE features remain unknown[13-15,23]. One possible source could be fullerene complexes with other astronomically abundant species. Already shortly after its discovery in 1985[24], the exceptional stability of $C_{60}$ led Kroto to speculate on its relevance to unidentified interstellar spectroscopic features, including the potential presence of exohedral charge transfer complexes of fullerenes with cosmically abundant metals, [$C_{60}$-Metal]$^+$ (ref.[25-27]). The metal atom will break the icosahedral symmetry of $C_{60}$ and the expected strong $C_{60}$-metal interactions will affect the vibronic energy levels, introducing many active vibrational modes, similar to those recently reported for $C_{60}H^+$ (ref.[28]), as well as a wealth of electronic transitions. However, due to a lack of accurate gas-phase spectroscopic data, no observational confirmation of [C$_{60}$-Metal]$^+$ complexes in space has been reported[23,27].

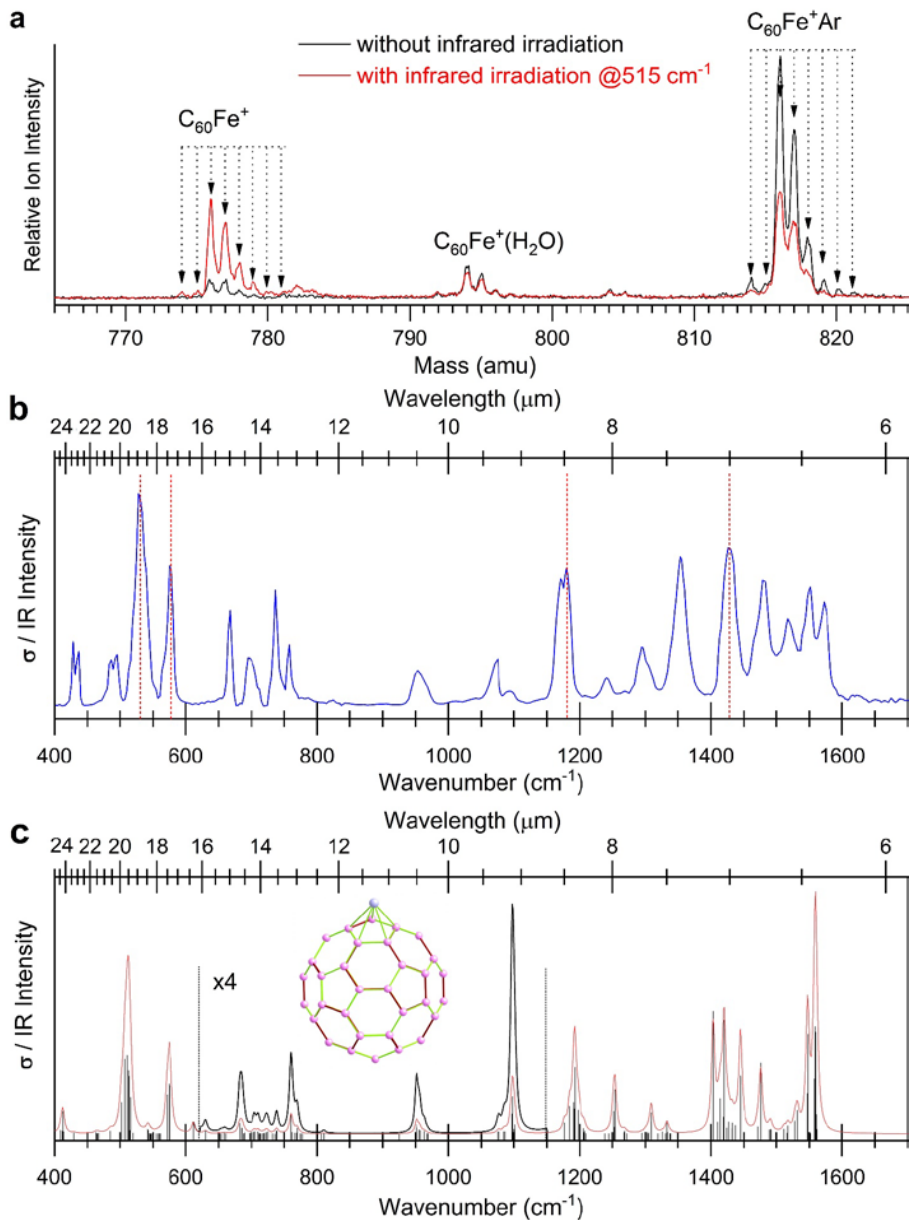

**Figure 1 | Laboratory synthesis and infrared spectroscopy of C$_{60}$Fe$^+$. a.** Mass distributions of C$_{60}$Fe$^+$ and its Ar-tagged complexes synthesized using a dual-target dual-laser vaporization source, with (red) and without (black) the infrared irradiation at 515 cm$^{-1}$. **b.** IRMPD spectrum of Ar-tagged C$_{60}$Fe$^+$ in the 400–1700 cm$^{-1}$ range. The positions of four neutral C$_{60}$ bands are indicated with vertical red dotted lines. **c.** Theoretically simulated spectrum of C$_{60}$Fe$^+$ at the BPW91/6-31G(d) level (red curve) with its calculated structure shown as inset. The calculated intensities are plotted with sticks and convolved using Lorentzian line shapes of 6 cm$^{-1}$ full width at half maximum. The convolved spectrum is enlarged by a factor of four between 650–1150 cm$^{-1}$ (black curve).

We developed an experimental protocol to measure, for the first time, the laboratory infrared (6–25 μm) spectra of [C$_{60}$-Metal]$^+$ (Metal = Fe and V) complexes via messenger-tagged infrared multiple photon dissociation (IRMPD) spectroscopy, in which either an argon atom or a D$_2$ molecule serves as a weakly bound spectator to probe the absorption of infrared light by [C$_{60}$-

Metal]$^+$ (see details in Methods section). Figure **1a** shows the mass distributions of $C_{60}Fe^+$ and its Ar-tagged complexes with (red) and without (black) infrared irradiation at 515 cm$^{-1}$ (19.4 μm). The carbon and iron isotope distributions can readily be recognized.

Figure **1b** presents the laboratory IRMPD spectrum of Ar-tagged $C_{60}Fe^+$. Apart from four features coinciding with the $C_{60}$ bands (indicated by the vertical red dotted lines), many more vibrational bands are visible. Figure **1c** shows the theoretically simulated infrared spectrum of $C_{60}Fe^+$ using density functional theory (DFT) calculations (see details in Methods section). The calculations essentially reproduce all observed infrared features, except that some intensities are slightly off. Such minor discrepancies are common when comparing DFT calculations with IRMPD experiments, where the multiple-photon excitation process may cause deviations from the calculated linear absorption intensities[28,29]. The overlap of the experimental infrared spectral features of $C_{60}$ and $C_{60}Fe^+$ may be due to the almost neutral character of $C_{60}$ in $C_{60}Fe^+$, as our calculations show $C_{60}$ carries only +0.125 $e$ charge from natural population analysis. In Extended Data Fig. 1, the laboratory IRMPD spectrum of $D_2$-tagged $C_{60}V^+$ and its comparison with both DFT calculations and that of $C_{60}Fe^+$ is provided, indicating that the spectral patterns of $C_{60}V^+$ and $C_{60}Fe^+$ have similar characteristic features (see detailed band positions in supplementary Table S1).

The overall agreement between the theoretical and laboratory spectra of both $C_{60}Fe^+$ and $C_{60}V^+$ provides confidence that we can reliably predict the infrared spectra of $C_{60}$ complexes with other cosmically abundant metals, such as Li, Na, K, Mg, Ca, and Al[27]. Calculations show they all have infrared spectral patterns similar to that of $C_{60}Fe^+$ or $C_{60}V^+$ (supplementary Figure S1). Vibrational analysis indicates that most features involve motions of the $C_{60}$ cage, perturbed by attaching metal atoms. Hence, the laboratory spectrum of either $C_{60}Fe^+$ or $C_{60}V^+$ can be considered as a model for various [$C_{60}$-Metal]$^+$ complexes.

To assess the plausibility of [$C_{60}$-Metal]$^+$ surviving under interstellar conditions, we used collision theory and the estimated 1% of cosmic carbon locked in $C_{60}$ for evolved stars (supplementary Table S4)[23,27]. With an assumed $2 \times 10^4$ cm$^{-3}$ hydrogen number density, [$C_{60}$-Metal]$^+$ formation rates are, depending on the specific metal, in the range of $1 \times 10^{-7}$ to $4 \times 10^{-2}$ yr$^{-1}$ ($C_{60}$ has a lifetime exceeding $10^8$ yrs in space; supplementary Table S5 and ref. 27). The thermal dissociation rates of [$C_{60}$-Metal]$^+$ complexes at typical temperatures below 300 K depend on their stabilities, and are for $C_{60}V^+$ and $C_{60}Fe^+$, with binding energies of 2.82 and 2.25 eV, respectively, significantly smaller than the formation rates (supplementary Table S5). The strong dependence of photodissociation on the ultraviolet radiation field and optical depth, and its competition with radiative decay (via visible to infrared photon emission) and photoionization to higher charge states make an accurate estimate of the photodissociation rate challenging (see supplementary information). However, our analysis still indicates that [$C_{60}$-Metal]$^+$ could plausibly form and survive in certain astrophysical environments.

Our laboratory infrared spectra can be used to discuss the potential spectral impact of the presence of [$C_{60}$-Metal]$^+$ in space. Co-existence of different [$C_{60}$-Metal]$^+$ complexes in the same object will slightly modify the band positions and intensities (supplementary Figure S1 and Table S1). Figure **2e-f** show the similarities between the laboratory $C_{60}Fe^+$ spectrum and a summed theoretical spectrum of [$C_{60}$-Metal]$^+$. The band shift at 18.9 μm is due to an underestimation of that mode frequency by DFT calculations, also seen in calculations for neutral $C_{60}$ (blue dashed lines in Figure **2f**; see also supplementary Figure S1). The overall agreement in band positions confirms that the laboratory spectra can serve as faithful models for various [$C_{60}$-Metal]$^+$ complexes.

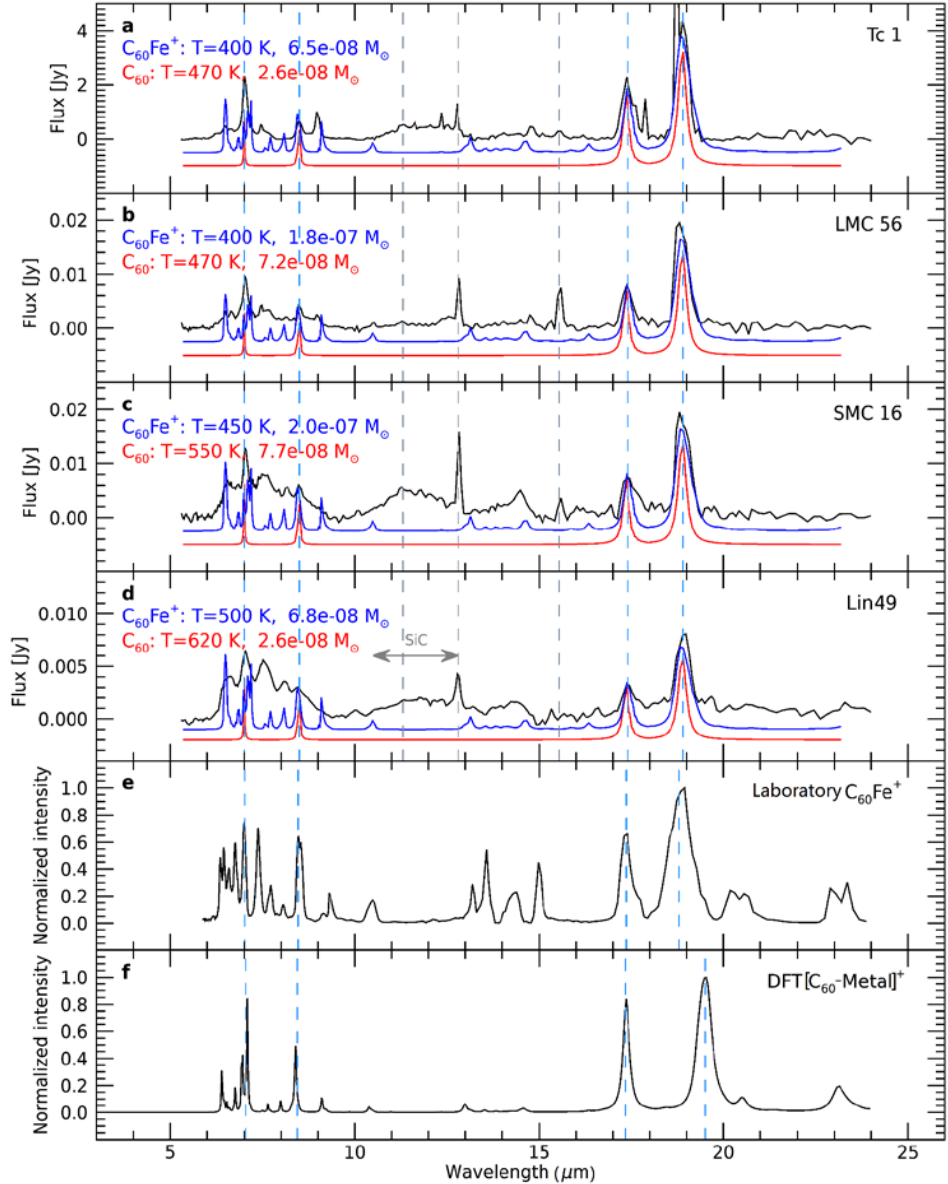

**Figure 2 | Comparison between the Spitzer infrared spectra of four fullerene-rich PNe (black) and the thermal emission model for C$_{60}$ (red) and C$_{60}$Fe$^+$ (blue). a.** Tc 1, **b.** LMC 56, **c.** SMC 16, **d.** Lin49. The spectra of Tc 1, LMC 56 and SMC 16, and Lin49 are adapted from Cami, et al. (ref. 13), Bernard-Salas, et al. (ref. 14), and Otsuka, et al. (ref. 15). The excitation temperatures and mass derived from the thermal emission models based on a microcanonical ensemble are indicated at the left in each panel. The models have taken into account the uncertainties of DFT predicted frequencies to align with the observation (see details in Methods section). Note that the 7.0 μm bands in Tc 1 and SMC 16 contain a 15% contribution of the [ArII] line (ref. 14). Blue and gray dashed lines indicate the four observational C$_{60}$ bands, and strong atomic emission lines, such as [NeII] (12.81 μm) and [NeIII] (15.55 μm) (ref. 13), respectively. The emission plateau between 11 and 13 μm (strong in Tc 1, SMC 16, and Lin49 but weaker in LMC 56) — indicated with a gray horizontal arrow — is generally attributed to and nominally labeled as SiC (ref. 14), although the different profiles and substructures of the plateau suggest other carriers may also be present (ref. 19). One possibility is that they may be due to the collective out-of-plane bending modes emitted by a mixture of aliphatic side groups attached to aromatic rings (ref. 30). **e.** Laboratory infrared spectrum of C$_{60}$Fe$^+$. **f.** Summed theoretical spectrum of [C$_{60}$-Metal]$^+$. Here, blue dashed lines indicate the DFT calculated frequencies of the four neutral C$_{60}$ bands.

Figure **2a-d** displays the continuum subtracted Spitzer infrared spectra of Tc 1, LMC 56, SMC 16, and Lin49 (refs.[13-15]), with observational and laboratory band positions summarized in supplementary Table S1. Note that to ease the comparison, all the Spitzer data have been rebinned to the wavelength grid of SMC 16 using linear interpolation, which may smear out a few sharp atomic lines in particular for the high-resolution spectral region (9.8–19.5 μm) of the Spitzer data of Tc 1 (ref.[13]; see details in Methods section), but it will not affect the following analysis. The four characteristic features of neutral $C_{60}$ (vertical blue dashed lines) are clearly seen in all four PNe, but with varying intensity ratios, and are observed also in the laboratory infrared spectra of $C_{60}Fe^+$ and $C_{60}V^+$ within the experimental and observational uncertainty. The spread in intensity ratios was reported also in many other $C_{60}$-containing astrophysical environments[14,15,19,52]. It has been noted that the 17.4 and 18.9 μm features of LMC 56, SMC 16, and Lin49 have small shoulders at longer wavelengths, for example, shoulders at about 17.8 and 18.1 μm, and 20.2 and 20.6 μm, as well as asymmetric band profiles[14,15,19]. Such characteristics could indicate the presence of other contributing emitters[14], as recently proposed from two laboratory bands of $C_{60}H^+$ at 17.7 and 19.1 μm[28].

Here, we find that [$C_{60}$-Metal]$^+$ complexes could potentially a) regulate the ratios of the four $C_{60}$ bands by the variation of their spectral intensities for different metals (supplementary Figure S1); b) explain the shoulders and asymmetric band profiles observed in the 17–20 μm region; and c) explain several other yet unidentified features.

In the 17–20 μm region, the laboratory spectra of both $C_{60}Fe^+$ and $C_{60}V^+$ show two intense bands at 17.4 and 18.8 μm with smaller features at 17.8 and 18.1 μm. Although it should be noted that for $C_{60}V^+$ the laboratory feature at 18.8 μm partially overlaps with $D_2$-tag induced bands at 19.3 and 19.8 μm (asterisks in Extended Data Fig. 1**b**), hindering the identification of pure $C_{60}V^+$ features in the laboratory experiment, our calculations do show an asymmetric profile for the 19 μm band (Extended Data Fig. 1**c**). In addition, the laboratory spectral feature of $C_{60}Fe^+$ at 18.8 μm is also asymmetric with a small shoulder around 19.5 μm.

In the 10–17 μm region of the laboratory spectra of both $C_{60}Fe^+$ and $C_{60}V^+$, there are several well-resolved features. Most of them coincide with features in LMC 56 including features superimposed on the 11–13 μm plateau, and several with features in Tc 1, SMC 16, and Lin49, within 0.1–0.2 μm intervals (supplementary Table S1). Exceptions are that no clear indications are seen in the observations for the 13.5 and 14.3 μm features of $C_{60}Fe^+$. The small wavelength differences could well be accounted for, considering that even for the $C_{60}$ bands in Tc 1 there is about 1% deviation between laboratory and observational band positions[13]. Beside known atomic lines (supplementary Table S1), to date no confirmed carriers could explain the emission features in this spectral region, except for a possible contribution of $C_{60}^+$ to the 10.5 μm band[16]. The presence of $C_{70}$ and its contribution to emission features in those objects, except for Tc 1 (ref.[13]), were proposed but are uncertain[14,15,19].

Apart from the two features at 7.0 and 8.5 μm coinciding with neutral $C_{60}$ bands, our laboratory spectra of both $C_{60}Fe^+$ and $C_{60}V^+$ also show other clear features in the 6–10 μm region that could have observational counterparts within 0.1 μm (supplementary Table S1). The exception is the laboratory double-peak feature of $C_{60}V^+$ at 9.1 and 9.3 μm, which may be hidden in the broad emission plateau (particularly in SMC 16 and Lin49, see Extended Data Fig. 3). Although $C_{60}^+$ has been suggested to potentially contribute to UIE features at 6.4, 7.1, and 8.1 μm[16], it seems the observational emission bands, for example the asymmetric profile of the intense 7.0 μm band of LMC 56, are better explained by [$C_{60}$-Metal]$^+$ complexes. Brieva et al. previously tentatively assigned the feature at 6.49 μm to a combination mode of neutral $C_{60}$, but importantly they also proposed that it could be due to the formation of fullerene-metal complexes[31], which now gains experimental support from our laboratory experiments.

**Table 1 | Cross-correlation coefficients**. The calculated r values for $C_{60}Fe^+$, $C_{60}V^+$, $C_{60}H^+$, $C_{60}$, $C_{60}^+$, and $C_{60}C^+$ templates and the observational Spitzer PNe data of Tc 1, LMC 56, SMC 16, and Lin49.

| PNe | $C_{60}Fe^+$ | $C_{60}V^+$ | $C_{60}H^+$ | $C_{60}Fe^+$ | $C_{60}V^+$ | $C_{60}H^+$ | $C_{60}$ | $C_{60}^+$ | $C_{60}C^+$ |
|---|---|---|---|---|---|---|---|---|---|
| | Laboratory band position and DFT intensity | | | DFT band position and intensity | | | | | |
| Tc 1 | 0.65 | 0.75 | 0.47 | 0.67 | 0.76 | 0.53 | 0.82 | 0.38 | 0.55 |
| LMC 56 | 0.74 | 0.82 | 0.60 | 0.75 | 0.83 | 0.65 | 0.85 | 0.50 | 0.70 |
| SMC 16 | 0.67 | 0.72 | 0.68 | 0.72 | 0.70 | 0.70 | 0.66 | 0.64 | 0.67 |
| Lin49 | 0.68 | 0.70 | 0.78 | 0.74 | 0.69 | 0.80 | 0.56 | 0.74 | 0.82 |

To support this proposition, we quantify the similarity between each of the four PNe spectra and the laboratory $C_{60}Fe^+$ spectrum by the cross-correlation coefficient, r, of the observational data with a $C_{60}Fe^+$ template, and the template is constructed using laboratory measured band positions and DFT calculated intensities (see details in Methods section). Table 1 summarizes the calculated coefficients which are also indicated in Extended Data Fig. 2a-d. The same procedure is used to quantify the cross-correlation of the observational spectra and the laboratory spectra of $C_{60}V^+$ and $C_{60}H^+$ (Table 1 and Extended Data Figs. 3-4). We additionally calculated the cross-correlation coefficients for similar template functions of $C_{60}Fe^+$, $C_{60}V^+$, $C_{60}H^+$, $C_{60}$, and $C_{60}^+$ utilizing only DFT calculated quantities (positions and intensities, Table 1 and Extended Data Figs. 5-8). Those analyses show that $C_{60}$ and $C_{60}$-metal complexes have r values close to 0.70–0.80, indicating a strong positive linear relationship, while $C_{60}H^+$ and $C_{60}^+$ only have a moderate linear relationship with the Spitzer PNe spectra except for Lin49 (ref.[15]). Moreover, since $C_{60}$ may accrete by absorbing a carbon atom to form $C_{60}C^+$ (ref.[27]), similar as [$C_{60}$-Metal]$^+$ complexes, we also did the same type analysis for $C_{60}C^+$, and the results in Table 1 and Extended Data Figs. 5-8 show that the r values for $C_{60}C^+$ are not better than [$C_{60}$-Metal]$^+$ complexes except for Lin49. This provides statistical support for the appearance of [$C_{60}$-Metal]$^+$ complexes in the Spitzer data, at a similar or slightly higher level of statistical confidence than $C_{60}$, $C_{60}^+$, $C_{60}H^+$ in Tc 1, LMC 56, and SMC 16[28]. It is consistent with our estimates based on collision theory that [$C_{60}$-Metal]$^+$ complexes could form and survive in space.

Closer inspection of Figure 2 and Extended Data Figs. 2-3 allows to pinpoint the wavelength regions between 6–9 and 13–15 μm, and the asymmetric profiles in the 17–20 μm region as potentially diagnostic for the presence of [$C_{60}$-Metal]$^+$. The characteristic features are identified to be due to the presence of metal on the $C_{60}$ cage, which perturbs the vibronic energy levels of $C_{60}$ both geometrically and electronically (supplementary Figures S2-S4). Our laboratory spectra also show features at 22.9–23.4 μm for $C_{60}Fe^+$ and at 23.5 μm for $C_{60}V^+$, attributed to metal–$C_{60}$ cage stretching vibrations, which may have weak coinciding bands in LMC 56, SMC 16, and Lin 49. Similarly, the laboratory double-peak feature of $C_{60}Fe^+$ at 20.2 and 20.6 μm might be reflected in particularly SMC 16 and Lin 49 (highlighted in light yellow in Extended Data Fig. 2). However, the current quality of the observation and the experiment in this wavelength range precludes a firm identification. The exact spectral positions of such metal–cage stretching vibrations depend on the specific metals as shown by our DFT calculations (supplementary Figures S1 and S5), and could serve as potential signatures to characterize and quantify a specific $[C_{60}\text{-Metal}]^+$ complex. The metal–$C_{60}$ vibrational modes beyond 30 μm are even more metal-specific (supplementary Figures S1 and S3).

The experimental work in this contribution is limited to the mid- to far-infrared spectral range. At shorter wavelengths, ranging from near-infrared to ultraviolet, $[C_{60}\text{-Metal}]^+$ also offers a wealth of electronic transition lines that could be potential candidates for DIBs. Because no experimental work exists for $[C_{60}\text{-Metal}]^+$ in these spectral ranges, we only note that, in comparison to $C_{60}^+$, recently identified as carrier of several DIBs[20-22], for $[C_{60}\text{-Metal}]^+$ many more electronic transition lines are predicted in our time-dependent DFT (TDDFT) calculations (supplementary Figure S4). It should be pointed out that TDDFT is currently not capable to accurately predict the exact energies of the electronic absorption bands, and the lack of experimental benchmark data hampers a quantitative comparison between theory and the DIB observations.

The origin and formation mechanisms of fullerenes in space are still mysterious[23,32]. Recalling the catalytic roles of metals in the nucleation and growth of various carbon nanostructures, including carbon nanotubes[33] and carbon cages under oxygen- and hydrogen-rich conditions[34,35], we conjecture that the presence of fullerene-metal complexes can link the chemical pathways and mechanisms of the formation and evolution of various carbonaceous species and thus carbon chemistry in space. Furthermore, one may even conjecture that a fraction of identified $C_{60}$ fullerene lines actually may stem from metal-fullerene complexes. To substantiate this argument, we have determined the temperature and abundance required for each species to replicate the observed emission spectra of the four fullerene-rich PNe using thermal emission models that treat the fullerene-metal complexes as a microcanonical ensemble (Figure 2). About $2.6$–$7.7\times10^{-8}$ solar mass ($M_\odot$) of pure $C_{60}$ is required to reproduce the four main emission bands. An equally good fit, which also accounts for some additional emission features, is obtained for ~$0.69$–$2.1\times10^{-7}$ $M_\odot$ of $C_{60}Fe^+$, or ~$1.9$–$6.1\%$ of the available carbon with excitation temperatures above ~400 K. Arguably, these diffuse nebulae do not disfavour the ionized form of gaseous $C_{60}$ and $C_{60}Fe$ owing to their low ionization potentials, but it is challenging to accurately estimate the fractions considering the complicated charge balance, i.e., competition among ionization, electron attachment, and recombination events as well as the lack of parameters associated with those processes (see details in Methods section). This argument together with the high cross-correlation coefficients and good fits of the thermal emission model supports the inference that gaseous ionized fullerene-metal complexes are potentially good candidates for explaining the infrared emission spectra of these four PNe in addition to neutral $C_{60}$. The forthcoming high resolution and high sensitivity data from the James Webb Space Telescope can provide better constraints on the environmental conditions of specific fullerene-metal complexes and their abundances, of crucial importance for revealing the dominant chemical pathways, and in particular the initial steps of fullerene formation in the Universe.

## Methods

**Experimental**. The [$C_{60}$-Metal]$^+$ (Metal = Fe and V) complexes are synthesized in vacuum by a dual-target dual-laser source, which is a modified version of the Smalley-type laser vaporization source used in the discovery of $C_{60}$ in 1985 (ref.[24]), and has been described in detail previously[36]. The choice of iron and vanadium as representative examples in the experiment is motivated by the fact that i) both iron and vanadium have been detected in many astrophysical environments[27,37,38], and ii) the synthesis of $C_{60}Fe^+$ and $C_{60}V^+$ is stable, facilitating the infrared experiment. Both the bulk metal and the $C_{60}$-fullerene targets are vaporized by 532 nm laser pulses from two independent Nd:YAG lasers, both operated at 10 Hz repetition rates. The vaporized neutral $C_{60}$ molecules and metal plasma, containing metal cations, collide with each other in the presence of He gas, introduced through a pulsed valve with a 6 bar stagnation pressure, which triggers formation and cooling of the complexes. We assume [$C_{60}$-Metal]$^+$ complexes to be thermalized to room temperature before expansion into vacuum, moderately cooling their internal degrees of freedom.

The high stabilities of $C_{60}Fe^+$ and $C_{60}V^+$, characterized by the 2.25 and 2.82 eV binding energies between $C_{60}$ and $Fe^+$ and $V^+$, respectively, (corresponding to approximately 20 photons at 10 μm) prevent the recording of high-quality infrared (IR) spectra over the full wavelength range. Only a few IR features of large calculated IR oscillator strengths show up with a low signal-to-noise (S/N) ratio (see for example the $C_{60}V^+$ case in supplementary Figure S6). Therefore, the messenger tagging technique was utilized, with Ar as tag for $C_{60}Fe^+$ and $D_2$ for $C_{60}V^+$. This technique has been well established in the past decades to obtain the vibrational or electronic spectra of molecular species via photodissociation[20,39-41]. Ar-tagged $C_{60}Fe^+$ (calculated binding energy is negligible) was formed by seeding about 5% Ar into the He carrier gas; for $D_2$-tagged $C_{60}V^+$ (0.3–0.4 eV, or 3 photons at 10 μm) a 0.5% $D_2$ seeding was used. After expansion into vacuum through a conical nozzle, the cluster beam is formed and shaped by a 2 mm diameter skimmer and a 2 mm slit aperture, before entering into the extraction zone of a perpendicular reflectron time-of-flight (TOF) mass spectrometer.

IR multiple photon dissociation (IRMPD) experiments are performed by overlapping the shaped cluster beam with IR light of the Free Electron Laser for Intra-Cavity Experiments FELICE[39]. The measurements were conducted with two FEL settings, covering the 400–850 and 700–1700 cm$^{-1}$ spectral ranges, at a repetition rate of 5 Hz. This allowed to record successive mass spectra with and without IR laser interaction. Laser excitation in resonance with a vibrational mode heats up the clusters by multiple photon absorption and intra-molecular vibrational redistribution (IVR). When the internal energy of the cluster is high enough, fragmentation (dissociation) takes place on the time scale of the experiment via the lowest-energy fragmentation channel, in the current case desorption of Ar from $C_{60}Fe^+\cdot Ar$, or of $D_2$ from $C_{60}V^+\cdot(D_2)_{1,2}$. Here, the Ar atom and $D_2$ molecule serve as non-interfering weakly bound messengers that will be shed if IR radiation is resonantly absorbed, representing a sensitive probe by mass spectrometry. Care has been taken with the FEL light focusing to minimize saturation effects and to avoid destruction of the $C_{60}$ cage, which is confirmed by the absence of $C_{60-2m}^+$ fragments.

The FEL was scanned in wavenumber steps of 5 cm$^{-1}$, which implies a non-constant step size in wavelength of 0.03–0.05 μm below 10 μm (above 1000 cm$^{-1}$), 0.1 μm around 800 cm$^{-1}$ (12.5 μm), and 0.2 μm around 500 cm$^{-1}$ (20 μm). The IR wavelength is calibrated using a grating spectrometer, with a typical uncertainty of the FEL wavelength of 0.2%.

The experiments allow to obtain IR spectra by comparing the mass spectrometric intensities of [$C_{60}$-Metal]$^+$ and its messenger-tagged complexes with [$I(\nu)$] and without ($I_0$) FEL light irradiation. The IRMPD yield is defined as the depletion ratio of $I(\nu)/I_0$,

$$Y(\nu) = I(\nu)/I_0.$$

(1)

To reduce the noise originating from cluster synthesis fluctuations, we first calculate the branching ratio $B$ of the number of $C_{60}Fe^+Ar_1$ ions to all $C_{60}Fe^+Ar_{0,1}$ ions, or $C_{60}V^+(D_2)_2$ ions to all $C_{60}V^+(D_2)_{0-2}$ ions

$$B = I[C_{60}Fe^+Ar_1] / \sum I[C_{60}Fe^+Ar_{0,1}].$$

(2)

$$B = I[C_{60}V^+(D_2)_2] \: / \sum I[C_{60}V^+(D_2)_{0,1,2}]. \tag{3}$$

Under the assumption of constant Ar or $D_2$ adsorption rates, this eliminates fluctuations in the synthesis. We then calculate the depletion yield $Y'$ as a function of IR frequency $v$ by taking the natural logarithmic ratio of the branching ratios with and without IR irradiation,

$$Y'(v) = -\ln[B(v)/B_0]. \tag{4}$$

The depletion yield $Y'(v)$ is divided by the laser pulse energy $E(v)$ to account for variation of the laser power (but note that this will not correct the intensity issues intrinsic to the multiphoton process) and to approximate the infrared absorption cross section,

$$\sigma(v) = Y'(v)/E(v). \tag{5}$$

The FEL laser pulse energy is reconstructed by measuring the pulse energy of a fraction of the pulse that is outcoupled of the FELICE cavity[42]. Typical pulse energies used range from 1000 mJ at 400 cm$^{-1}$ to 200 mJ at 1600 cm$^{-1}$.

To confirm there are no other dissociation channels in the $C_{60}V^+$ experiment apart from $D_2$ desorption, the laser induced changes of the ion intensity and the summed ion intensity of $C_{60}V^+(D_2)_{0\text{-}2}$ with and without IR light irradiation are plotted in supplementary Figure S7 as a function of IR wavelength. It is obvious that the vast majority of the gained $C_{60}V^+$ signal arises from $C_{60}V^+(D_2)_2$ depletion, while the summed ion intensity is conserved within the experimental uncertainty. A similar procedure has been conducted for the experiments with $C_{60}Fe^+$.

Without $D_2$ tagging, only a few features of high calculated oscillator strengths are observed in the infrared spectrum of $C_{60}V^+$ in the 1100–1600 cm$^{-1}$ range (supplementary Figure S6). These indicate that the perturbation of the $D_2$-tag on the vibrational band positions of $C_{60}V^+$ is negligible, which is confirmed by density functional theory (DFT) calculations on the infrared spectra of $C_{60}V^+$ and its $D_2$-tagged complexes in the 400–1700 cm$^{-1}$ range (supplementary Figure S8). The calculations show that the main changes induced by $D_2$-tag are that strong vibrational features around 510, 540, and 830 cm$^{-1}$, which significantly involve $D_2$ motions, show up (supplementary Figure S9). For the experiments on $C_{60}Fe^+$, no much influence of Ar has been found in the studied spectral range. Hence, the IRMPD spectra of messenger-tagged $C_{60}V^+$ and $C_{60}Fe^+$ can be regarded as the IR absorption spectra of $C_{60}V^+$ and $C_{60}Fe^+$.

**Theoretical**. For $C_{60}V^+$, five structures, i.e., $\eta^5$, $\eta^6$, $\eta^{2(6\text{-}6)}$, $\eta^{2(6\text{-}5)}$, and $\eta^1$ have been fully optimized to obtain its most stable structure. In those structures, the V atom binds with the hexagonal center, pentagonal center, bridge of a hexagon and a hexagon, bridge of a hexagon and a pentagon, and atop of $C_{60}$, respectively. Since the electron configuration of the V atom is [Ar]3d$^3$4s$^2$, and the ground state of V$^+$ is [Ar]3d$^3$4s$^1$, the interaction of V$^+$ with the closed-shell molecule $C_{60}$ to form $C_{60}V^+$ could result in S = 0, 1, 2 spin states. Hence, for each structure, these spin states were considered. During the structure optimization, the $\eta^1$ structure was unstable and converged to either $\eta^5$ or $\eta^6$ in all three spin states. Three functionals, i.e., BPW91, PBE, and B3LYP, were employed to test the reliability of DFT calculations. All three functionals calculate the $\eta^5$ structure in S = 2 spin state to be most stable (supplementary Table S2). Similar calculations have been performed for $C_{60}Fe^+$, for which the $\eta^5$ structure with S = 5/2 was the most stable. From the comparison in supplementary Figure S10, it can be seen that overall BPW91 provides the best agreement between the calculated IR spectrum and experiment, consistent with previous findings on the good performance of this functional in calculating vanadium-carbon binary clusters[43]. Due to the large size of the studied complex, double-$\zeta$ quality basis sets (6-31G(d) and/or def2-SVP) were used for all calculations. Triple-$\zeta$ quality basis sets (6-311G(d), def2-TZVP, and Lanl2TZ) were also checked, showing consistent results with the double-$\zeta$ basis sets (supplementary Figure S11). Therefore, the BPW91/6-31G(d) method was employed to do the calculations for $C_{60}$, $C_{60}^+$, and [$C_{60}$-Metal]$^+$ (Metal = H, Li, Na, K, Mg, Ca, Al, and Fe) as well as for $C_{60}C^+$. The harmonic approximation was employed to perform the vibrational analysis of the calculated structures.

The vibrational analysis is used on one hand to confirm that the calculated structures are real minima and on the other hand to simulate theoretical IR spectra to compare with the measured IRMPD spectra. All calculations were conducted with Gaussian09 program package[44].

**Details on the Astronomical Data and the Cross-Correlation Analysis.** The Spitzer infrared spectra of Tc 1, LMC 56 and SMC16, and Lin49 are adapted from Cami, et al. (ref. 13), Bernard-Salas, et al. (ref. 14), and Otsuka, et al. (ref. 15). They were all recorded by the Infrared Spectrograph (IRS) on board the Spitzer Space Telescope. For Tc 1, the entire spectrum consists of observations at low resolution mode ($\lambda/\Delta\lambda \sim 60$–120) with the Short-Low module (SL, 5.2–14 μm; $\Delta\lambda \sim 0.04$–0.12 μm), and at high-resolution mode ($\lambda/\Delta\lambda \sim 600$) with the Short-High (SH, 9.8–19.5 μm; $\Delta\lambda \sim 0.02$–0.04 μm) and Long-High (LH, 19.5–36 μm; $\Delta\lambda \sim 0.04$–0.06 μm) modules. For LMC 56, SMC 16, and Lin49, all spectra were recorded at low resolution using the SL and Long-Low (LL) modules. In the current work, all Spitzer data have been rebinned to the wavelength grid of SMC 16 using linear interpolation for the ease of comparison. Note this rebining process may smear out a few sharp atomic lines in particular for the high-resolution spectral region (9.8–19.5 μm) of the Spitzer data of Tc 1 (ref. 13), but it will not affect the following analysis. For more details about those observational data, please refer to refs. 13-15.

The Spitzer data of the four PNe (Tc 1, LMC 56, SMC16, and Lin49) have a high signal-to-noise ratio with typical root mean square (RMS) value of ~1% below ~14 μm, and ~5–20% for ~14–25 μm. However, the spectral resolution of the Spitzer data is not high enough, except for Tc 1 in the 9.8–36 μm spectral range, to spectrally resolve individual molecular emission lines. Hence, the largest uncertainty for the detection of spectral fingerprints is steered by uncertainties in the continuum subtraction[13] and the fact that other molecules and dust species contribute to the overall spectrum as well. That uncertainty, $\sigma_{cont}$, is quantified by the standard deviation of the flux in the wavelength region between 20.1–21.7 μm. For a molecule with a number of $N$ spectral fingerprints in a particular wavelength region, the $3\sigma$ detection limit can be estimated as $3\sigma_{cont}/\sqrt{N}$. In Extended data Figs. 2-8, the thickness of the horizontal, light gray lines indicates the $3\sigma$ detection limit for the fingerprints of the metal complexes.

The similarity between each of the four Spitzer PNe spectra and laboratory infrared spectra is quantified by the cross-correlation coefficient, which is a preferred and well-established method of choice for the analysis of one component in a complex system[45]. For instance, it has been used for detecting chemical species in exoplanet atmospheres[46, 47] or for the study of $C_{60}^+$ diffuse interstellar band correlations and environment variations[48]. Here it is [$C_{60}$-Metal]$^+$ in the PNe with varying chemical compositions and thermodynamic properties. The cross-correlation coefficient of two sample populations $X$ and $Y$ as a function of the lag ($L$) is calculated as

$$r(L) = \begin{cases} \frac{\sum_{k=0}^{N-|L|-1}(x_{k+|L|}-\bar{x})(y_k-\bar{y})}{\sqrt{\left[\sum_{k=0}^{N-1}(x_k-\bar{x})^2\right]\left[\sum_{k=0}^{N-1}(y_k-\bar{y})^2\right]}} & \text{for } L < 0 \\ \frac{\sum_{k=0}^{N-L-1}(x_k-\bar{x})(y_{k+L}-\bar{y})}{\sqrt{\left[\sum_{k=0}^{N-1}(x_k-\bar{x})^2\right]\left[\sum_{k=0}^{N-1}(y_k-\bar{y})^2\right]}} & \text{for } L \geq 0 \end{cases}. \tag{6}$$

where $\bar{x}$ and $\bar{y}$ are the means of the sample populations $x = (x_0, x_1, \cdots\cdots, x_{N-1})$ and $y = (y_0, y_1, \cdots\cdots, y_{N-1})$.

However, a direct calculation of the cross-correlation coefficient between observational Spitzer spectra and the laboratory spectra is not meaningful owing to i) the multiphoton nature of the laboratory experiment that results in the experimental intensities not being a good measure of the line strength, and in the case of $C_{60}V^+$, ii) a few features significantly involve $D_2$ motions. These shortcomings can be overcome by using the outcome of DFT calculations, in particular the calculated line strengths. We recalibrated the laboratory spectra by taking the experimental band positions (but excluding the bands strongly involving $D_2$ motions for $C_{60}V^+$) and scaling the intensities to DFT calculations. In this step, we account for the laboratory spectral resolution determined by the scan step, $\sigma_{lab}$, or, if DFT calculated frequencies are used, the ~2% uncertainty in DFT calculated band positions[43]. The outcome is then convolved with a Gaussian function with the full width at half maximum being determined from the Spitzer spectral resolution, $\sigma_{obs}$. Finally, we assign the combined laboratory (or DFT) and Spitzer wavelength uncertainty, $\sqrt{\sigma_{lab}^2 + \sigma_{obs}^2}$, as wavelength resolution to the new $C_{60}Fe^+$ or $C_{60}V^+$ template (red curves in Extended data Figs. 2-8) and calculate the cross-correlation coefficient of this template with the PNe spectra; see Table 1. Note that since we allow the frequencies to move around given the uncertainty in lab/DFT frequencies and observational wavelength grid, the templates shown in Extended data Figs. 2-8 are slightly different from one source to the other. The entire spectra were employed for the cross-correlation analysis.

**Abundance estimates.** The Spitzer spectra of the four PNe allow for the abundance of the contribution species to be calculated. Since both $C_{60}$ and $C_{60}Fe^+$ share the same most intense bands at 7.0, 8.5, 17.4, and 18.9 μm, a reliable relative abundance estimate of one species with respect to the other is not possible. However, one can estimate the abundance of one species so that the predicted band intensities reproduce the observed spectra. We therefore follow the same methodology as Cami et al.[13] by assuming a microcanonical thermal emission model. Instead of relying on experimentally obtained relative absorption coefficients for the $C_{60}$ bands, we here exploit the DFT calculated band intensities. For a thermal population distribution over the excited states, we determine the excitation temperature and total mass of each species following a $\chi^2$ minimization procedure for the four main bands at 7.0, 8.5, 17.4, and 18.9 μm (Figure 2). We here account for the described deviation, which allows a shift around the calculated frequencies of maximally ~2% (ref. [49]). For $C_{60}$, the excitation temperature ranges between ~470–620 K, while the excitation temperature of $C_{60}Fe^+$ is lower for all targets and ranges between ~400–500 K. About $2.6 - 7.7 \times 10^{-8} M_\odot$ of pure $C_{60}$ and $0.69 - 2.1 \times 10^{-7} M_\odot$ of $C_{60}Fe^+$ are required to reproduce the observed emission bands. At the measured carbon abundance for PNe in the galaxy, LMC, and SMC[50], and assuming a mass-loss rate of $10^{-4}$ $M_\odot yr^{-1}$ over the past 100 years[13], this implies that $C_{60}$ represents ~0.7–2.3% of the available carbon, consistent with previous estimates for evolved stars in supplementary Table S4. Under the assumption that the emission bands are purely due to $C_{60}Fe^+$, ~1.9–6.1% of the available carbon would be consumed by $C_{60}Fe^+$.

Cami et al. derived a thermal excitation temperature of $C_{60}$ in Tc 1 of 330 K[13], slightly lower than our value of 470 K. That low temperature allowed Cami et al. to argue that the emission does not originate from free molecules in the gas phase, but from $C_{60}$ formed on the surface of cool, solid material with most likely composition being carbonaceous grains. In that case, the charge effects on individual molecules is negligible. For a typical effective temperature of ~20,000 K the radiation field peaks for photon energies in the range of 6 to 10 eV. Hence, in contrast to solid fullerenes, gaseous $C_{60}$ may be largely ionized owing to its low ionization potential of 7.61 eV[51]. If these large gaseous clusters are free-flying species, temperatures above ~450 K can be readily reached via stochastic heating by absorption of a single UV photon[52]. The excitation temperature of $C_{60}$ is highly uncertain owing to uncertainties in the Einstein A coefficients[53], but most reported $C_{60}$ sources have excitation temperatures well above 400 K, even reaching values of 1000 K, which may question the existence of $C_{60}$ in solid form. Our simplified thermal modelling approach supports these higher excitation temperatures, and hence supports the argument of the presence of gaseous fullerene-rich species. Taking the argument one step further, if $C_{60}$ is in a gaseous form, a significant fraction could be ionized. Hence, neutral $C_{60}$, possessing only four active infrared features, alone cannot explain the infrared emission features here under study. However, it is challenging to accurately estimate the fractions considering the complicated charge balance, i.e., competition among ionization, electron attachment, and recombination events as well as the lack of parameters associated with those processes. The high correlation coefficients (see Table 1), the good fits of the thermal emission model (Figure 2), and the low ionization potential of $C_{60}Fe$ (ca. 6.34 eV) render the conjecture that gaseous ionized fullerene-metal complexes, in which $C_{60}$ behaves as almost neutral, are equally good candidates as $C_{60}$ for explaining the infrared spectra of the four PNe. However, more accurate Einstein A coefficients are required to draw solid conclusions.

## Data availability

The data that support the findings of this study are available from the corresponding author on reasonable request. Any methods, additional references and data are available in supplementary information.

**Acknowledgements** This work is supported by the KU Leuven Research Council (C1 grants C14/18/073 and MAESTRO C16/17/007), the Research Foundation Flanders (AKUL/15/16 G0H1416N and G0A0519N), and the project CALIPSOplus under the Grant Agreement 730872 from the EU Framework Programme for Research and Innovation HORIZON 2020. L.D. acknowledges support from the ERC-CoG grant 646758. The computational resources and services used in this work were provided by the VSC (Flemish Supercomputer Center), funded by the Research Foundation–Flanders (FWO) and the Flemish Government–department EWI. We gratefully acknowledge the Nederlandse Organisatie voor Wetenschappelijk Onderzoek (NWO) for the support of the FELIX Laboratory and thank the FELIX staff. G.L.H. also acknowledges the support of Xi'an Jiaotong University via the "Young Talent Support Plan" and the "Fundamental Research Funds for Central Universities", and thanks Profs. Yong Zhang (Sun Yat-sen University) and Hong Gao (Chinese Academy of Sciences) for helpful discussions, and Dr. Alessandra Candian (University of Amsterdam) for providing the Spitzer data files in Figure 2. We also thank Prof. Alexander Tielens (Leiden University) for useful suggestions, and Dr. Jan Vanbuel and Dr. Piero Ferrari for assistance in adapting the laser vaporization source with the molecular beam setup at FELICE.

**Author contributions** G.L.H. conceived and coordinated the work, performed the IRMPD experiments with O.V.L. and J.M.B., formulated the idea of this work from discussion with E.J., P.L., and L.D., conducted the theoretical calculations and modelling. G.L.H. calculated the formation and dissociation rates, L.D. did the cross-correlation and abundance analyses, and J.M.B. did the normal mode decomposition analysis. G.L.H. wrote the manuscript with comments and inputs from E.J., L.D., J.M.B., and P.L. E.J. and P.L. obtained the funding for this research.

**Competing interests** The authors declare no competing interests.

**Correspondence and requests for materials** should be addressed gaolei.hou@xjtu.edu.cn (G.-L.H.) or leen.decin@kuleuven.be (L.D.)

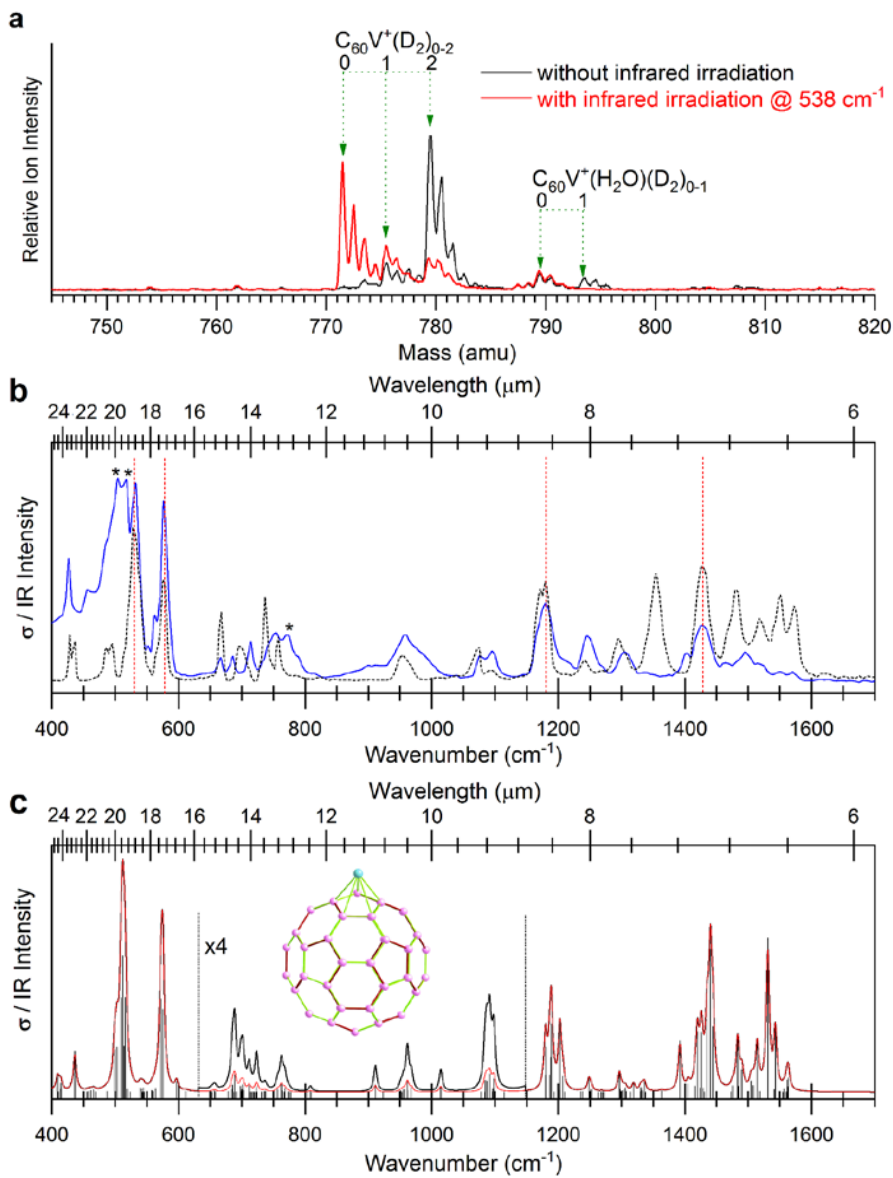

**Extended Data Fig. 1 | Laboratory synthesis and infrared spectroscopy of $C_{60}V^+$. a.** Mass distributions of $C_{60}V^+$ and its $D_2$-tagged complexes synthesized using a dual-target, dual-laser vaporization source, with (red) and without (black) the infrared irradiation at 538 cm⁻¹. **b.** IRMPD spectrum of $D_2$-tagged $C_{60}V^+$ (solid blue curve) and its comparison with that of Ar-tagged $C_{60}Fe^+$ (dotted black curve) in the 400–1700 cm⁻¹ range. The positions of four neutral $C_{60}$ bands are indicated with vertical red dotted lines; the asterisks mark the features mainly induced by $D_2$ (supplementary Figures S8-S9). **c.** Theoretically simulated spectrum of $C_{60}V^+$ at the BPW91/6-31G(d) level (red curve) with its calculated structure shown as inset. The calculated intensities are plotted with sticks and convolved using Lorentzian line shapes of 6 cm⁻¹ full width at half maximum. The convolved spectrum is also enlarged by a factor of four between 650–1150 cm⁻¹ (black curve). The comparison shows that the spectral patterns of $C_{60}Fe^+$ and $C_{60}V^+$ have similar characteristic features except that the spectrum of $C_{60}V^+$ show several $D_2$-involved modes and their intensities vary, likely due to multiple-photon excitation effects (see detailed band positions in supplementary Table S1).

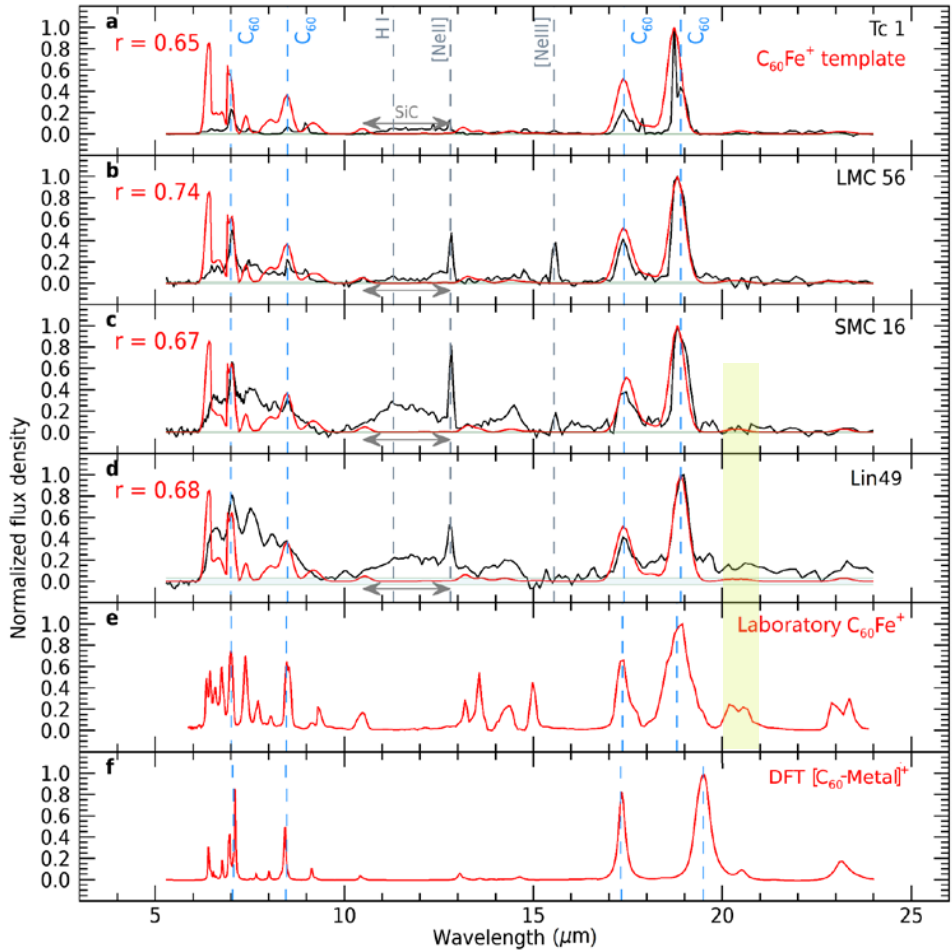

**Extended Data Fig. 2 | Comparison between the Spitzer infrared spectra of four fullerene-rich PNe (black) and the C$_{60}$Fe$^+$ template constructed by scaling the laboratory C$_{60}$Fe$^+$ spectrum to the DFT calculated band intensities (red). a.** Tc 1, **b.** LMC 56, **c.** SMC 16, **d.** Lin49. The spectra of Tc 1, LMC 56 and SMC 16, and Lin49 are adapted from Cami, et al. (ref. 13), Bernard-Salas, et al. (ref. 14), and Otsuka, et al. (ref. 15). The cross-correlation coefficients of the observational PNe spectrum and the C$_{60}$Fe$^+$ template are indicated in each panel. The thickness of the horizontal, light gray line indicates the 3σ detection limit (see details in Methods section). **e.** Laboratory infrared spectrum of C$_{60}$Fe$^+$. **f.** Summed theoretical spectrum of [C$_{60}$-Metal]$^+$. See caption of Figure 2 for more details.

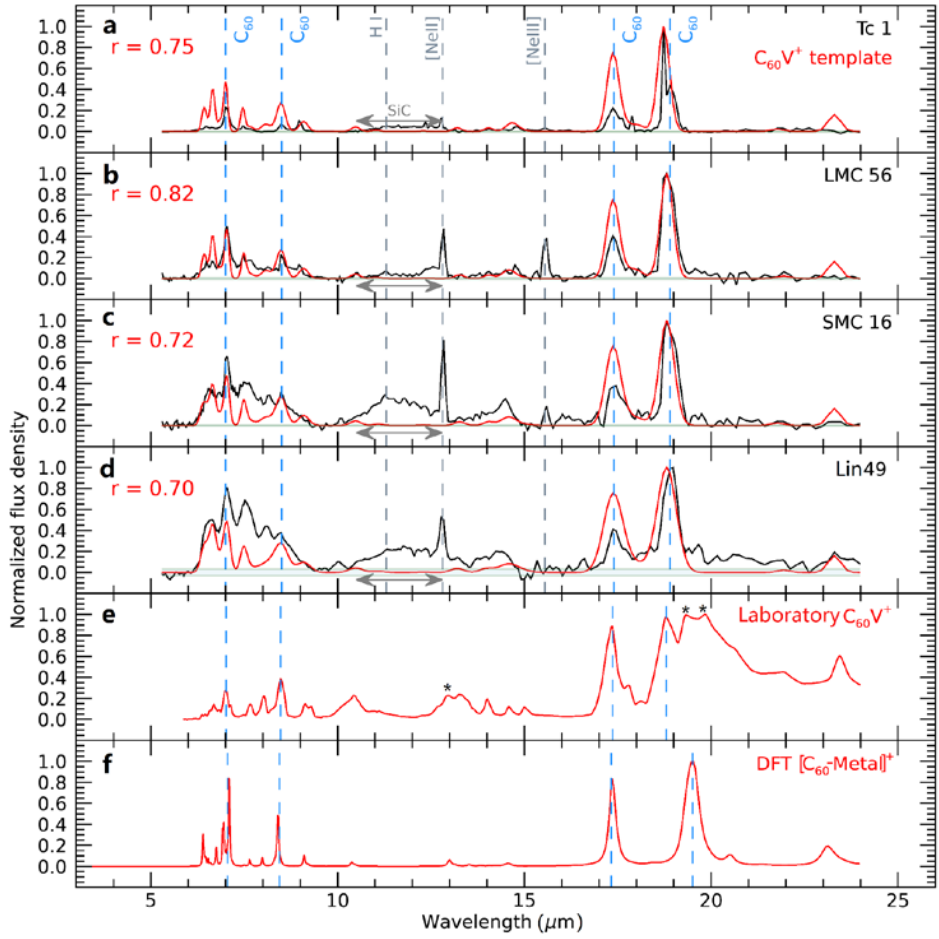

**Extended Data Fig. 3 | Comparison between the Spitzer infrared spectra of four fullerene-rich PNe (black) and the C₆₀V⁺ template constructed by scaling the laboratory C₆₀V⁺ spectrum to the DFT calculated band intensities (red). a.** Tc 1, **b.** LMC 56, **c.** SMC 16, **d.** Lin49. The spectra of Tc 1, LMC 56 and SMC 16, and Lin49 are adapted from Cami, et al. (ref. 13), Bernard-Salas, et al. (ref. 14), and Otsuka, et al. (ref. 15). The cross-correlation coefficients of the observational PNe spectrum and the C₆₀V⁺ template are indicated in each panel. The thickness of the horizontal, light gray line indicates the 3σ detection limit (see details in Methods section). **e.** Laboratory infrared spectrum of C₆₀V⁺. The asterisks indicate D₂-tag induced bands, hence without possible UIE counterparts. **f.** Summed theoretical spectrum of [C₆₀-Metal]⁺. See caption of Figure 2 for more details.

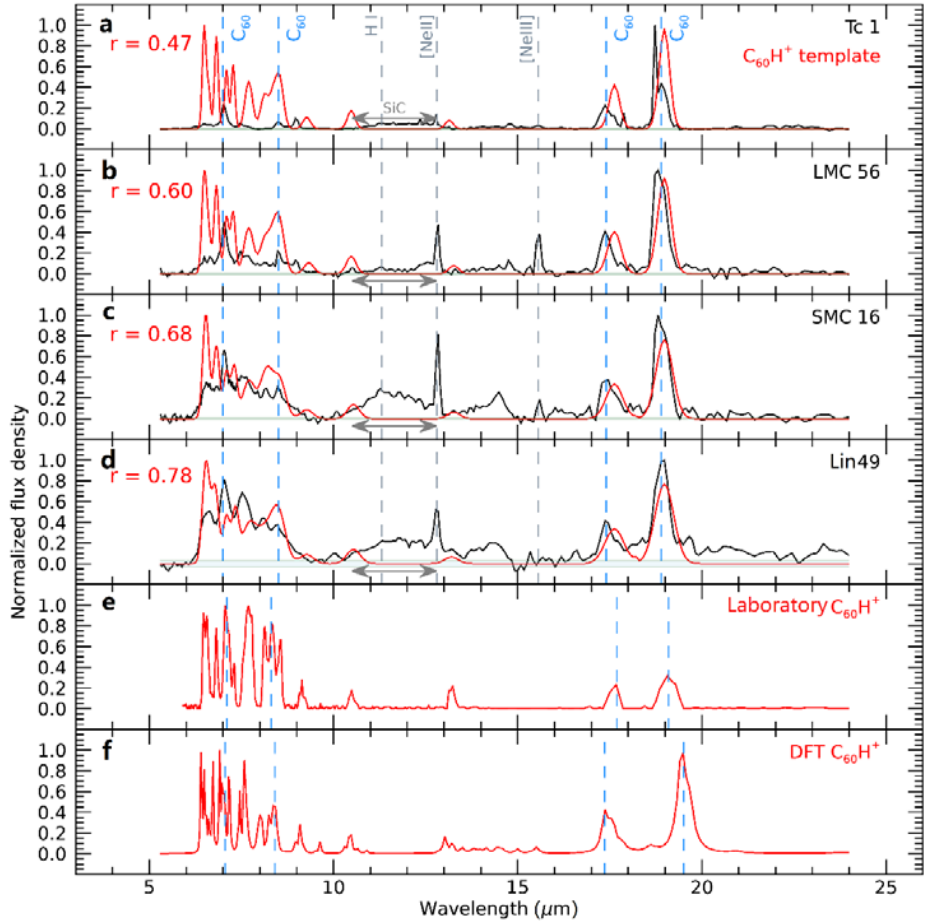

**Extended Data Fig. 4 | Comparison between the Spitzer infrared spectra of four fullerene-rich PNe (in black) and the $C_{60}H^+$ template constructed by scaling laboratory $C_{60}H^+$ spectrum to the DFT calculated intensities (in red). a.** Tc 1, **b.** LMC 56, **c.** SMC 16, **d.** Lin49. The spectra of Tc 1, LMC 56 and SMC 16, and Lin49 are adapted from Cami, et al. (ref. 13), Bernard-Salas, et al. (ref. 14), and Otsuka, et al. (ref. 15). The cross-correlation coefficients of the observational PNe spectrum and the $C_{60}H^+$ template are indicated in each panel. The thickness of the horizontal, light gray line indicates the 3σ detection limit (see details in Methods section). **e.** Laboratory infrared spectrum of $C_{60}H^+$ (ref. 28). **f.** BPW91/6-31G(d) calculated spectrum of $C_{60}H^+$. See caption of Figure 2 for more details.

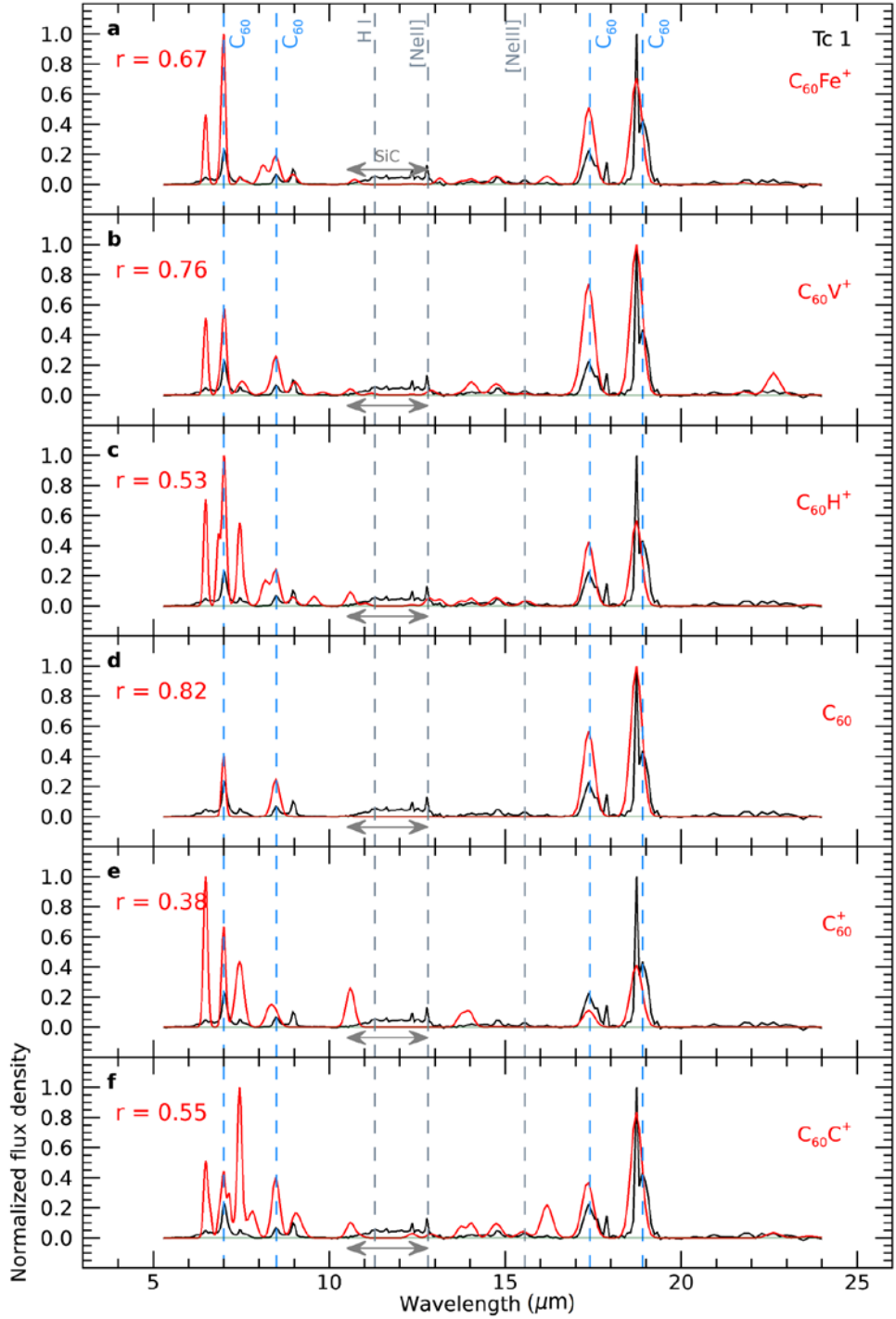

**Extended Data Fig. 5 | Comparison between the Spitzer infrared spectrum of Tc 1 (in black) and the DFT calculated spectra (in red). a.** $C_{60}Fe^+$, **b.** $C_{60}V^+$, **c.** $C_{60}H^+$, **d.** $C_{60}$, **e.** $C_{60}^+$, and **f.** $C_{60}C^+$. The cross-correlation coefficients of the observational PNe spectrum and the $C_{60}$ related species are indicated in each panel. The blue dashed lines indicate the four $C_{60}$ bands; gray dashed lines some of the strong atomic emission lines, for example, [NeII] (12.81 μm) and [NeIII] (15.55 μm) (ref. 13). The thickness of the horizontal, light gray line indicates the 3σ detection limit (see details in Methods section). See caption of Figure 2 for more details.

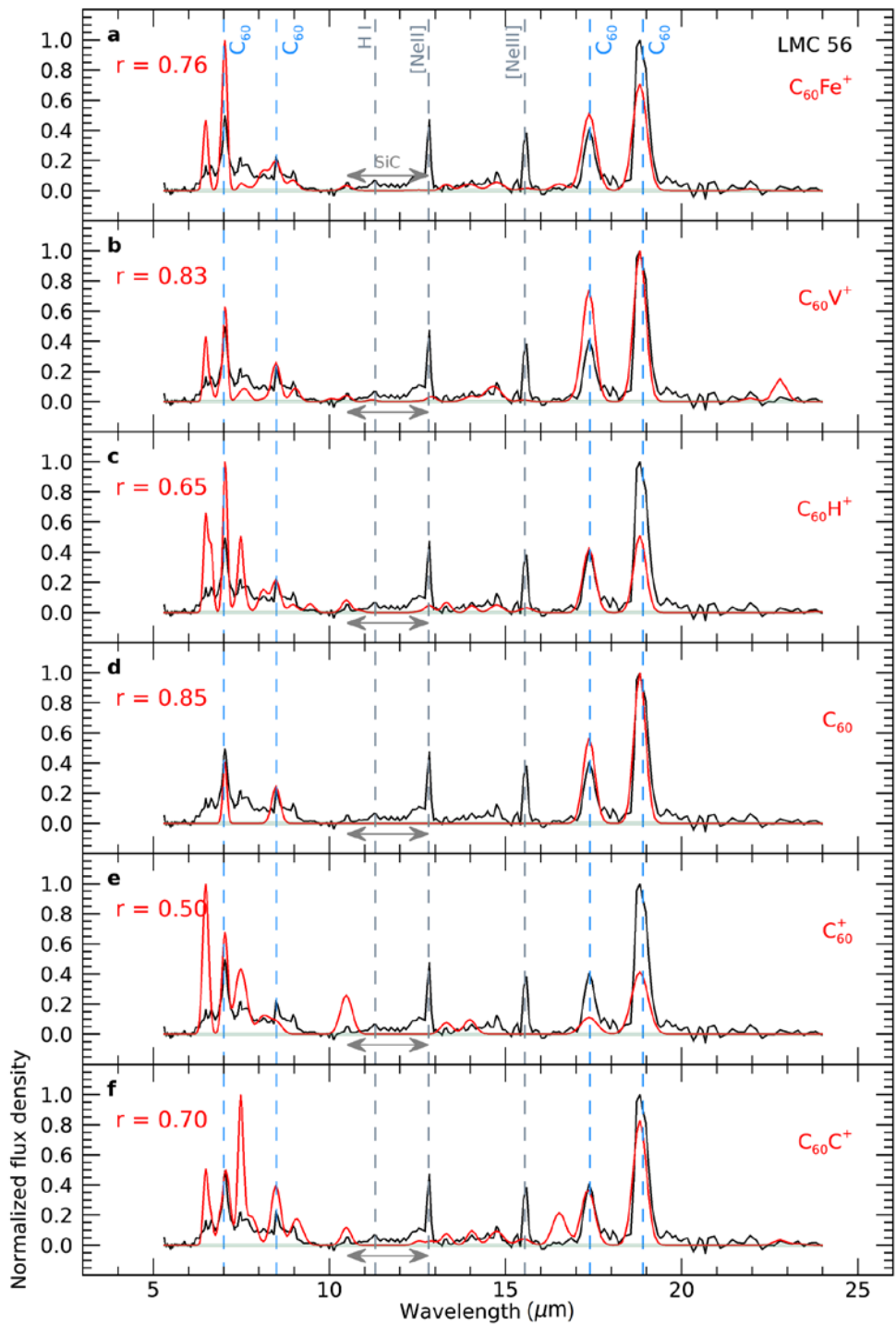

**Extended Data Fig. 6 | Comparison between the Spitzer infrared spectrum of LMC 56 (in black) and the DFT calculated spectra (in red). a.** $C_{60}V^+$, **b.** $C_{60}V^+$, **c.** $C_{60}H^+$, **d.** $C_{60}$, **e.** $C_{60}^+$, and **f.** $C_{60}C^+$. The cross-correlation coefficients of the observational spectrum and the $C_{60}$ related species are indicated in each panel. The blue dashed lines indicate the four $C_{60}$ bands; gray dashed lines some of the strong atomic emission lines, for example, [NeII] (12.81 μm) and [NeIII] (15.55 μm) (ref. 13).The thickness of the horizontal, light gray line indicates the 3σ detection limit (see details in Methods section). See caption of Figure 2 for more details.

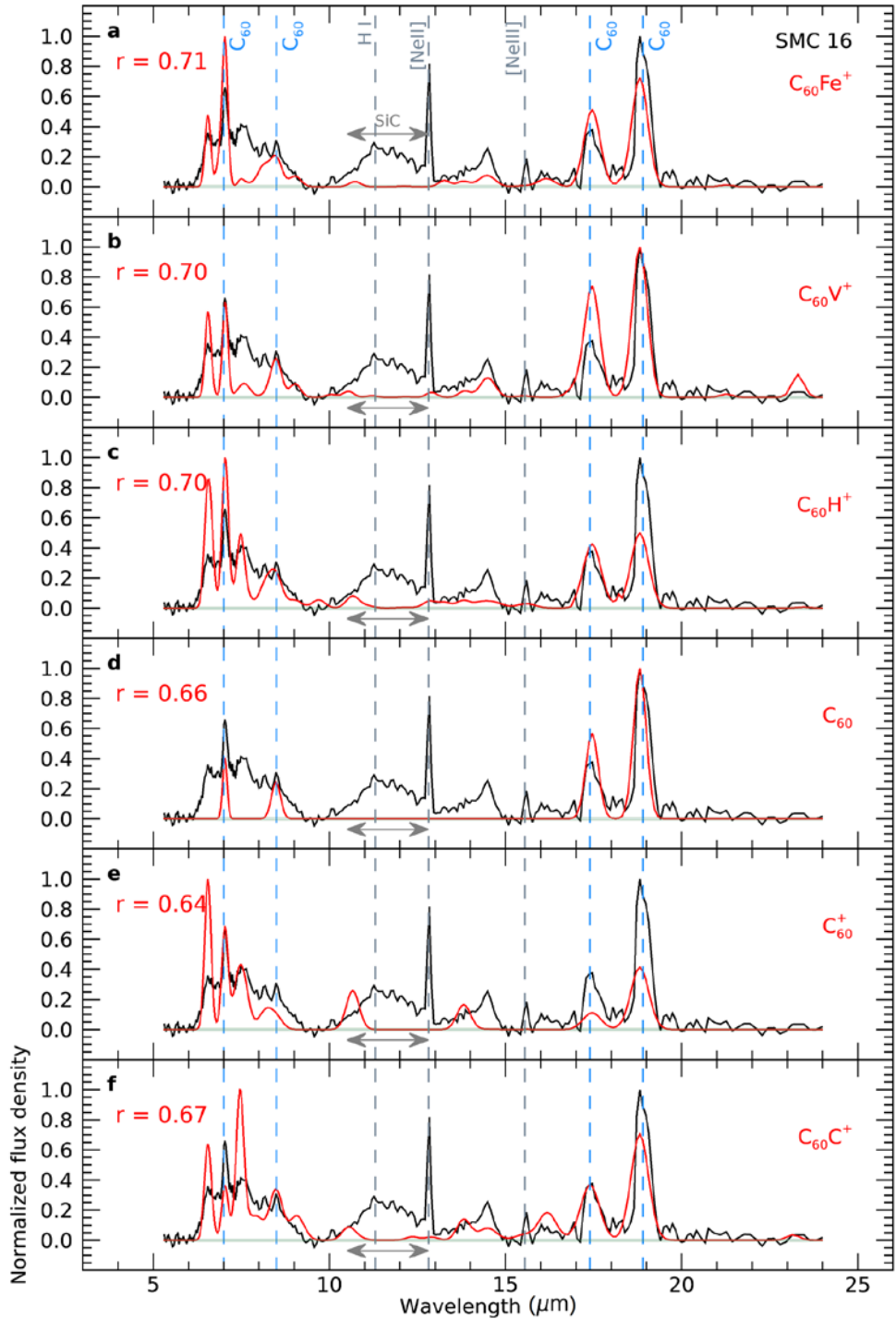

**Extended Data Fig. 7 | Comparison between the Spitzer infrared spectrum of SMC 16 (in black) and the DFT calculated spectra (in red). a.** $C_{60}Fe^+$, **b.** $C_{60}V^+$, **c.** $C_{60}H^+$, **d.** $C_{60}$, **e.** $C_{60}^+$, and **f.** $C_{60}C^+$. The cross-correlation coefficients of the observational spectrum and the $C_{60}$ related species are indicated in each panel. The blue dashed lines indicate the four $C_{60}$ bands; gray dashed lines some of the strong atomic emission lines, for example, [NeII] (12.81 μm) and [NeIII] (15.55 μm) (ref. 13).The thickness of the horizontal, light gray line indicates the 3σ detection limit (see details in Methods section). See caption of Figure 2 for more details.

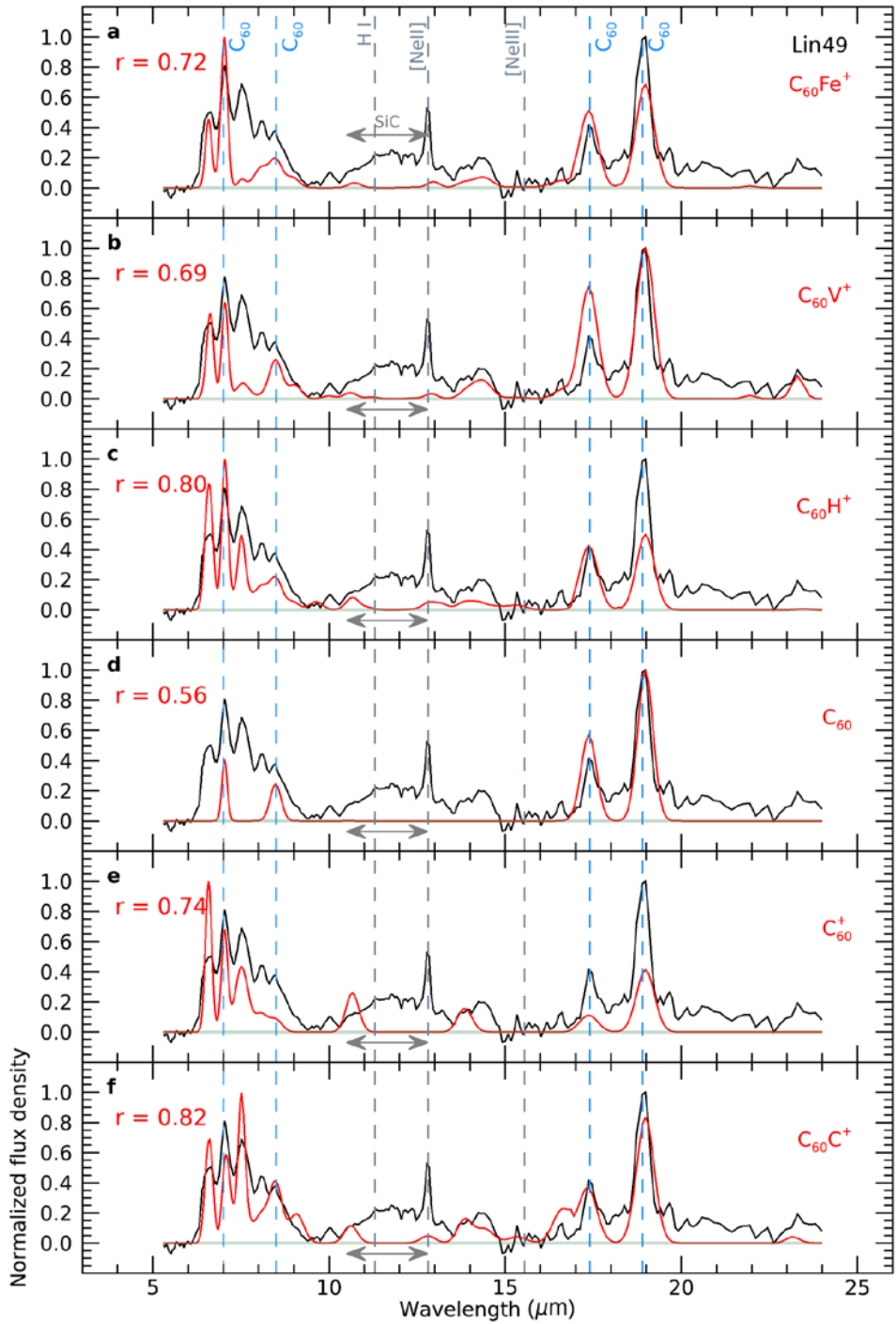

**Extended Data Fig. 8 | Comparison between the Spitzer infrared spectrum of Lin49 (in black) and the DFT calculated spectra (in red). a.** $C_{60}Fe^+$, **b.** $C_{60}V^+$, **c.** $C_{60}H^+$, **d.** $C_{60}$, **e.** $C_{60}^+$, and **f.** $C_{60}C^+$. The cross-correlation coefficients of the observational spectrum and the $C_{60}$ related species are indicated in each panel. The blue dashed lines indicate the four $C_{60}$ bands; gray dashed lines some of the strong atomic emission lines, for example, [NeII] (12.81 μm) and [NeIII] (15.55 μm) (ref. 13). The thickness of the horizontal, light gray line indicates the 3σ detection limit (see details in Methods section). See caption of Figure 2 for more details.

# Supplementary information

# Buckyball-metal complexes as promising carriers of astronomical unidentified infrared emission bands

Gao-Lei Hou[1,5]*, Olga V. Lushchikova[2], Joost M. Bakker[2], Peter Lievens[1], Leen Decin[3,4]*, Ewald Janssens[1]

[1]Quantum Solid-State Physics, KU Leuven, Celestijnenlaan 200D, 3001 Leuven, Belgium

[2]FELIX Laboratory, Radboud University, Toernooiveld 7, 6525 ED Nijmegen, The Netherlands

[3]Institute of Astronomy, KU Leuven, Celestijnenlaan 200D, 3001 Leuven, Belgium

[4]School of Chemistry, University of Leeds, Leeds, LS2 9JT, United Kingdom

[5]MOE Key Laboratory for Non-Equilibrium Synthesis and Modulation of Condensed Matter, School of Physics, Xi´an Jiaotong University, Xi´an, 710049 Shaanxi, China

Correspondence to: gaolei.hou@xjtu.edu.cn (GLH) or leen.decin@kuleuven.be (LD)

The supplementary information includes the formation and dissociation rates and additional figures and tables for the manuscript.

**Formation Rates of the [C₆₀-Metal]⁺ Complexes.** We employed collision theory to estimate the formation rates of [C₆₀-Metal]⁺ complexes in space. Collision theory predicts the rate $r(\text{T})$ of a bimolecular gas-phase reaction between reactants A and B at temperature T to be

$$r(\text{T}) = kn_\text{A}n_\text{B} = Z\rho \cdot \exp\left(\frac{-E_\text{a}}{RT}\right) \tag{S1}$$

where $k$ is the collisional rate constant in units of cm³s⁻¹, $n_\text{A}$ and $n_\text{B}$ the number densities of A and B (in cm⁻³), $Z$ the collision rate (in cm⁻³s⁻¹), $\rho$ the steric factor which reflects the reaction probability depending on the mutual orientations of the reactant particles, $E_\text{a}$ the activation energy of the reaction, and $R$ the ideal gas constant.

For two neutral particles, the collision rate can be approximated by a hard-sphere model,

$$Z = n_\text{A}n_\text{B}\sigma_\text{AB}\sqrt{\frac{8k_\text{B}T}{\pi\mu_\text{AB}}} \tag{S2}$$

where $\sigma_\text{AB} = \pi(r_\text{A} + r_\text{B})^2$ is the collision cross section with $r_\text{A}$ and $r_\text{B}$ the radii of A and B, $k_\text{B}$ the Boltzmann constant, and $\mu_\text{AB}$ the reduced mass of A and B. Assuming that there is no activation energy, i.e., $E_\text{a} = 0$, for the AB complex formation, and that the steric factor $\rho$ equals 1, the hard-sphere collisional rate constant can be simplified as

$$k_\text{HS} = \sigma_\text{AB}\sqrt{\frac{8k_\text{B}T}{\pi\mu_\text{AB}}} = 4.65 \times 10^{-12}(r_\text{A} + r_\text{B})^2\sqrt{\frac{T}{\mu_\text{AB}}} \tag{S3}$$

with $r_\text{A}$ and $r_\text{B}$ in Å, $T$ in K, $\mu_\text{AB}$ in atomic mass units, and $k_\text{HS}$ in of cm³s⁻¹.

For a collision between an ion and a small neutral particle without activation energy ($E_\text{a} = 0$), the rate constant $k_\text{L}$ is described by the Langevin model,

$$k_\text{L} = 2\pi e\sqrt{\frac{\alpha}{\mu_\text{AB}}} = 2.34 \times 10^{-9}\sqrt{\frac{\alpha}{\mu_\text{AB}}} \tag{S4}$$

where $e$ is the elementary charge in cgs units (statcoulombs), $\alpha$ the polarizability of the neutral particle in cgs units (Å³), and $k_\text{L}$ in units of cm³s⁻¹. The polarizability of C₆₀, $\alpha \approx 80$ Å³, and the radius of C₆₀ is 5.09 Å. The radii of the ionized metals are slightly smaller than their atomic radii.

Supplementary Table S3 lists the abundances of metals that we investigated in this work, and supplementary Table S4 summarizes the abundances of fullerenes by percent of gas-phase carbon locked in fullerene species in different astrophysical environments. Depending on the sources, the metal abundances are in the range of ~$(5 \times 10^{-11}$–$3.2 \times 10^{-5}) \cdot n_\text{H}$ with $n_\text{H}$ the number density of atomic hydrogen. The temperature in the planetary nebulae, in the interstellar medium in general, can vary significantly, from several tens to several thousands of Kelvin. Previously, using the thermal excitation model, Cami et al. derived a temperature range of 319–347 K for C₆₀ and 142–243 K for C₇₀ through fitting their observational band intensities in Tc 1 (ref. [1]). To obtain a wide range dependence on the temperature, we perform our calculations at two different temperatures of 50 and 300 K.

In space, the [C₆₀-Metal]⁺ complexes can form via i) C₆₀ + Metal⁺, ii) C₆₀⁺ + Metal, or iii) C₆₀ + Metal + hν (photoionization). From $k_\text{L1}$ (C₆₀ + Metal⁺) and $k_\text{L2}$ (C₆₀⁺ + Metal) in supplementary Table S5, it can be seen that the first two routes have comparable collision rate constants independent of temperature. Even ignoring the requirement of post ultraviolet (UV) photoionization after forming the neutral [C₆₀-Metal] complex, the neutral route ($k_\text{HS}$ in supplementary Table S5) has collision rate constants about one order of magnitude lower than the two ionic routes. It is challenging to estimate the photoionization efficiency for forming the [C₆₀-Metal]⁺ complexes via the neutral route as it depends on both the radiation flux and the photoionization cross sections of neutral [C₆₀-Metal] that are unknown. The abundances of Metal⁺ ions are also challenging to quantify. Hence, we simplify the problem by considering only the $C_{60}^+$ + Metal route to estimate a lower limit of the [$C_{60}$-Metal]$^+$ formation rate using $k_{L2}$, which is in the range of 1–4×10$^{-9}$ cm$^3$·s$^{-1}$ depending on the specific metals (supplementary Table S5). Hence, the formation rate of [$C_{60}$-Metal]$^+$ complexes per $C_{60}^+$ molecule is in the range of ~$(1.5 \times 10^{-19} - 6.3 \times 10^{-14}) \cdot n_H$ cm$^3$s$^{-1}$. Using a previously quoted value of $n_H < 2 \times 10^4$ cm$^{-3}$ by Omont in estimating the formation of $C_{60}H^+$ (ref.[2]), the [$C_{60}$-Metal]$^+$ formation rate is estimated to be $< 3 \times 10^{-15} - 1.2 \times 10^{-9}$ s$^{-1}$ (or $< 1 \times 10^{-7} - 4 \times 10^{-2}$ yr$^{-1}$) per $C_{60}^+$ molecule (supplementary Table S5).

The $C_{60}$ fullerene has a longer lifetime than the few $10^8$ yrs estimated lifetime of PAHs, from which it is thought to form[3]. In evolved stars (supplementary Table S4), around 1% of cosmic carbon is estimated to be locked in $C_{60}^+$, and the carbon abundance (supplementary Table S3) is about $(1.6–3.8 \times 10^{-4}) \cdot n_H$, which gives a number density of $C_{60}^+$ around 0.05–0.13 cm$^{-3}$ for $n_H < 2 \times 10^4$ cm$^{-3}$ and an estimated formation rate of [$C_{60}$-Metal]$^+$ complexes of $5 \times 10^{-9} - 5.2 \times 10^{-3}$ cm$^{-3}$yr$^{-1}$.

We note that this estimated formation rate may underestimate their abundance in space, because it assumes that the fullerenes already form and undergo random collisions with metal atoms to form the complexes. It is well-recognized that metals can catalyze the nucleation and growth of various carbon nanostructures, including carbon nanotubes[4] and carbon cages even under oxygen- and hydrogen-rich conditions[5,6]. We therefore conjecture that metals could play a role in the initial steps of forming various carbonaceous species including carbon cages and their complexes, enhancing their abundances in space. This mechanism can potentially link the chemical pathways and mechanisms of the formation and evolution of various carbonaceous species. For example, Ziurys and co-workers recently found that $C_{60}$ could form from the decomposition of SiC dust grains under interstellar relevant conditions[7]. It is known that SiC grains are commonly associated with objects where $C_{60}$ has been identified and that they contain various metals[1,7-9], making it plausible that the presence of metals could facilitate the decomposition of SiC and the formation of carbon cages, consistent with previous findings[5]. This potential formation pathway involving metals may help to explain the abundance of $C_{60}$ in hydrogen-rich environments, and could greatly enhance the abundance of fullerene-metal complexes in space, representing an interesting mechanism deserving further investigation in laboratory simulated astrophysical conditions.

**Thermal Dissociation Rates of the [$C_{60}$-Metal]$^+$ Complexes.** We approximate the unimolecular thermal dissociation rate ($k_d$) of the [$C_{60}$-Metal]$^+$ complex with an Arrhenius form,

$$k_d = A \cdot \exp\left(\frac{-E_a}{RT}\right) \tag{S5}$$

where $A$ is the prefactor, $R$ the ideal gas constant, $T$ the temperature of the complex, and $E_a$ the activation energy for which we use the calculated [$C_{60}$-Metal]$^+$ binding energy. To have a non-zero dissociation rate, the internal energy should be much higher than the activation energy (or binding energy) due to the intramolecular energy redistribution and the finite heat capacity of the complex. For example, the dominant dissociation channel of $C_{60} \rightarrow C_2 + C_{58}$ has an activation energy of 10.8±0.3 eV, but internal energies of about 50 eV are needed for dissociation on a timescale of tens to hundreds of microseconds[10]. Considering the similar binding strength and the same size of $C_{60}C^+$ with [$C_{60}$-Metal]$^+$, we here take the prefactor $A = 1.6 \times 10^{15}$ s$^{-1}$ determined previously for $C_{60}C^+$ to estimate their thermal dissociation rates[11,12]. In supplementary Table S5, we summarize the thermal dissociation rates at two temperatures of 50 and 300 K.

**Photodissociation Rates of the [$C_{60}$-Metal]$^+$ Complexes.** The complexes can also be destroyed by collision-induced dissociation or UV photodissociation. Collision-induced dissociation is not important due to the high binding energies of the complexes and the relatively low collision energies given the generally low temperature of the environment. The rate of UV photodissociation is challenging to estimate.

UV photodissociation implies dissociation of $[C_{60}\text{-Metal}]^+$ following the absorption of a UV photon. Since the photoabsorption rate is much lower than the vibrational cooling rate (see below), sequential absorption of UV photons can be ignored. Under these conditions, the UV photodissociation rate ($k_{PD}$) can be approximated by the product of the photoabsorption rate ($k_{PA}$) with the probability that the complex will dissociate if a single UV photon is absorbed ($p_d$).

Only photons with an energy between 6 and 9 eV are considered relevant, since competing decay rates are much higher than dissociation rates if the complex has internal energies below 6 eV (see below) and photoionization (to higher charge states) dominates upon absorption of photons with energies above 9 eV. The calculated ionization energies of $[C_{60}\text{-Metal}]^+$ (8.65, 8.95, and 9.04 eV for Ca, Fe, and V, respectively) are much lower than that of $C_{60}^+$ (18.98 ± 0.35 eV)[13]. The typical time scale for photoionization, which is an electronic process, is much shorter than that of dissociation, which involves atomic motion. In a Habing field, there are of the order of $10^8$ photons cm$^{-2}$ s$^{-1}$ between 6 and 9 eV[14]. The $C_{60}$ photoabsorption cross section in this photon energy range is about 200 Mb (1 Mb = 1× $10^{-18}$ cm$^2$)[15], which yields a $k_{PA}$ on the order of $2×10^{-8}$ s$^{-1}$. In photodissociation region (PDR) the number of photons can be a factor of $10^4$ higher, which yields an upper limit for $k_{PA}$ of $10^{-4}\text{s}^{-1}$.

Once a UV photon is absorbed and the internal energy of $[C_{60}\text{-Metal}]^+$ is higher than its binding energy, the complex can dissociate. The rate will, however, strongly depend on the internal energy, the density of states, the dissociation pathway, and the dissociation threshold of the complex. In earlier work, with a statistical model provided for $C_{60}H_2$, which has a binding energy of 2.9 eV comparable to that of $C_{60}V^+$, the dissociation rate $k_d$ was estimated to be on the order of $10^{-5}$–$10^{-1}$ s$^{-1}$ following absorption of a photon between 6 and 9 eV[14]. For $C_{60}V^+$, this rate is likely lower considering that the dissociation coordinate (V–$C_{60}$ cage stretching vibration; ~430 cm$^{-1}$) of $C_{60}V^+$ has a much lower frequency than that of $C_{60}H_2$ (H–$C_{60}$ cage stretching vibration; ~2800 cm$^{-1}$).

The dissociation process competes with other decay channels, greatly affecting the dissociation probability. The two most important competing decay channels are i) de-excitation via vibrational cooling and infrared emission and ii) de-excitation via recurrent fluorescence, which is likely important since $[C_{60}\text{-Metal}]^+$ has a dense spectrum of low-energy electronic transitions (supplementary Figure S4). Thermal population of low-energy electronic excited states leads to photon emission, so-called recurrent fluorescence, thereby cooling $[C_{60}\text{-Metal}]^+$ (ref.[16]). Taking into account the competing channels, the probability that the excited complex decays by dissociation is

$$p_d = \frac{k_d}{k_d + k_{VC} + k_{RF}} \tag{S6}$$

where $k_d$ is the dissociation rate as noted above, $k_{VC}$ is the vibrational cooling rate which has a typical value of 10–100 s$^{-1}$, and $k_{RF}$ is the rate of recurrent fluorescence from electronic excited states, which value is not known but typical values of $10^4$–$10^5$ s$^{-1}$ are found for metal, carbon or carbon-containing clusters[16]. That gives a value for $p_d$ on the order of $10^{-10}$–$10^{-5}$.

Combining the photoabsorption rate in the 6–9 eV range with the probability for dissociation, we obtain for the photodissociation rate $k_{PD} = k_{PA} \cdot p_d$ values on the order of $10^{-14} - 10^{-9}\text{s}^{-1}$or $10^{-6} - 10^{-1} \text{ yr}^{-1}$, comparable or slightly higher than the formation rate depending on different metals.

It should be noted that the photoabsorption cross section of $[C_{60}\text{-Metal}]^+$ is unknown and it can be different from that of $C_{60}$. The arguments point out that conducting an more accurate estimate than the above on the photodissociation rates of $[C_{60}\text{-Metal}]^+$ complexes is currently not possible.

From supplementary Table S5, it can be seen that the formation rates for the complexes with Metal = V and Fe are significantly larger than their thermal dissociation rates. Hence, considering both the thermal dissociation rates and the current challenges for estimating the accurate photodissociation rates given the above arguments, it is plausible that the $[C_{60}\text{-Metal}]^+$ complexes could form and survive in certain astrophysical environments. In particular, the complexes with high binding energies such as $C_{60}V^+$ and $C_{60}Fe^+$ could be able to survive.

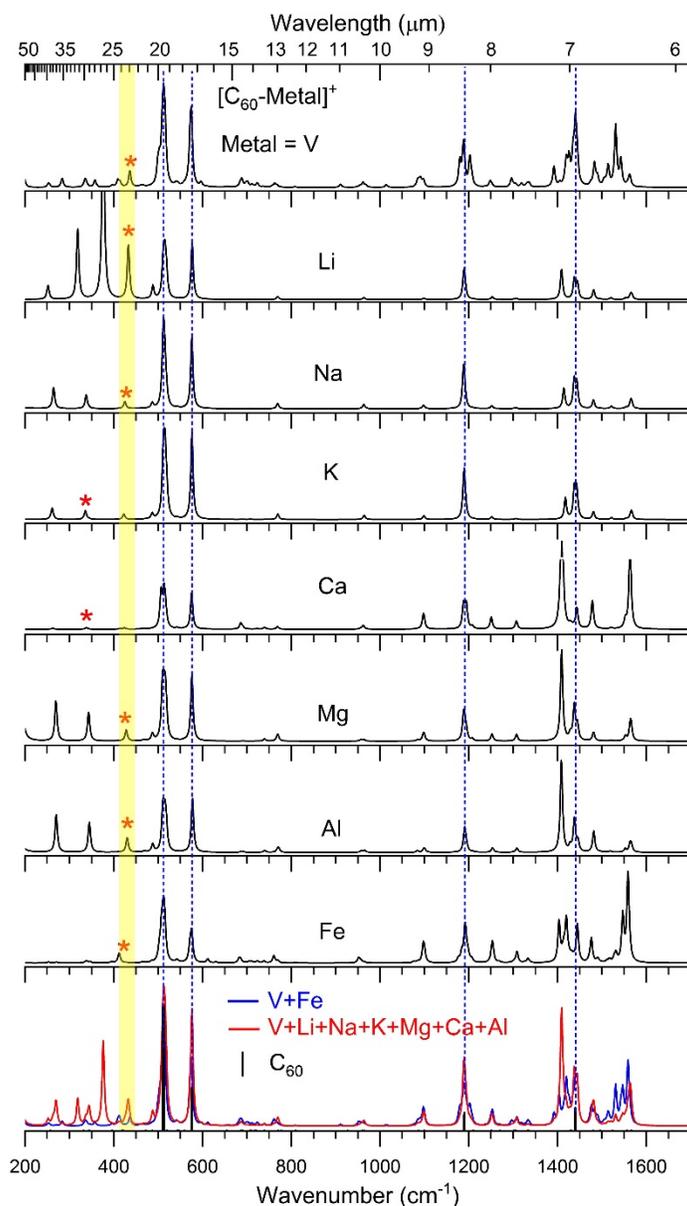

**Supplementary Figure S1 | Simulated infrared spectra of the most stable [$C_{60}$-Metal]$^+$ (Metal = V, Li, Na, K, Mg, Ca, Al, and Fe) and the stick spectrum showing the four neutral $C_{60}$ bands at BPW91/6-31G(d) level of theory.** The blue dashed lines show the coinciding features of [$C_{60}$-Metal]$^+$ with the four $C_{60}$ bands but with different intensity ratios. The spectra are convolved using Lorentzian line shapes of 6 cm$^{-1}$ full width at half maximum. They all show similar infrared spectral patterns regarding both the band positions and relative intensities. The small differences are due to different metals perturbing $C_{60}$ at different extents both geometrically and electronically. The average spectra for all metals and for vanadium and iron are also provided, and it can be seen that the averaged spectra of [$C_{60}$-Metal]$^+$ are similar to that of $C_{60}V^+$, except that the intensities of a few features, for example those at 680, 1100, 1420, 1540 cm$^{-1}$, vary and slightly shift within a few wavenumbers. Such comparison supports the idea to utilize the experimental infrared spectrum of $C_{60}V^+$ as a representative mode of various [$C_{60}$-Metal]$^+$ complexes to discuss their infrared spectral features. The asterisks (also highlighted with the yellow bar) indicate the vibrational modes due to metal–$C_{60}$ cage stretching, and their vectors are presented in supplementary Figure S2. Note that the $C_{60}K^+$ and $C_{60}Ca^+$ complexes have the metal-$C_{60}$ stretching modes at lower frequencies (~340 cm$^{-1}$).

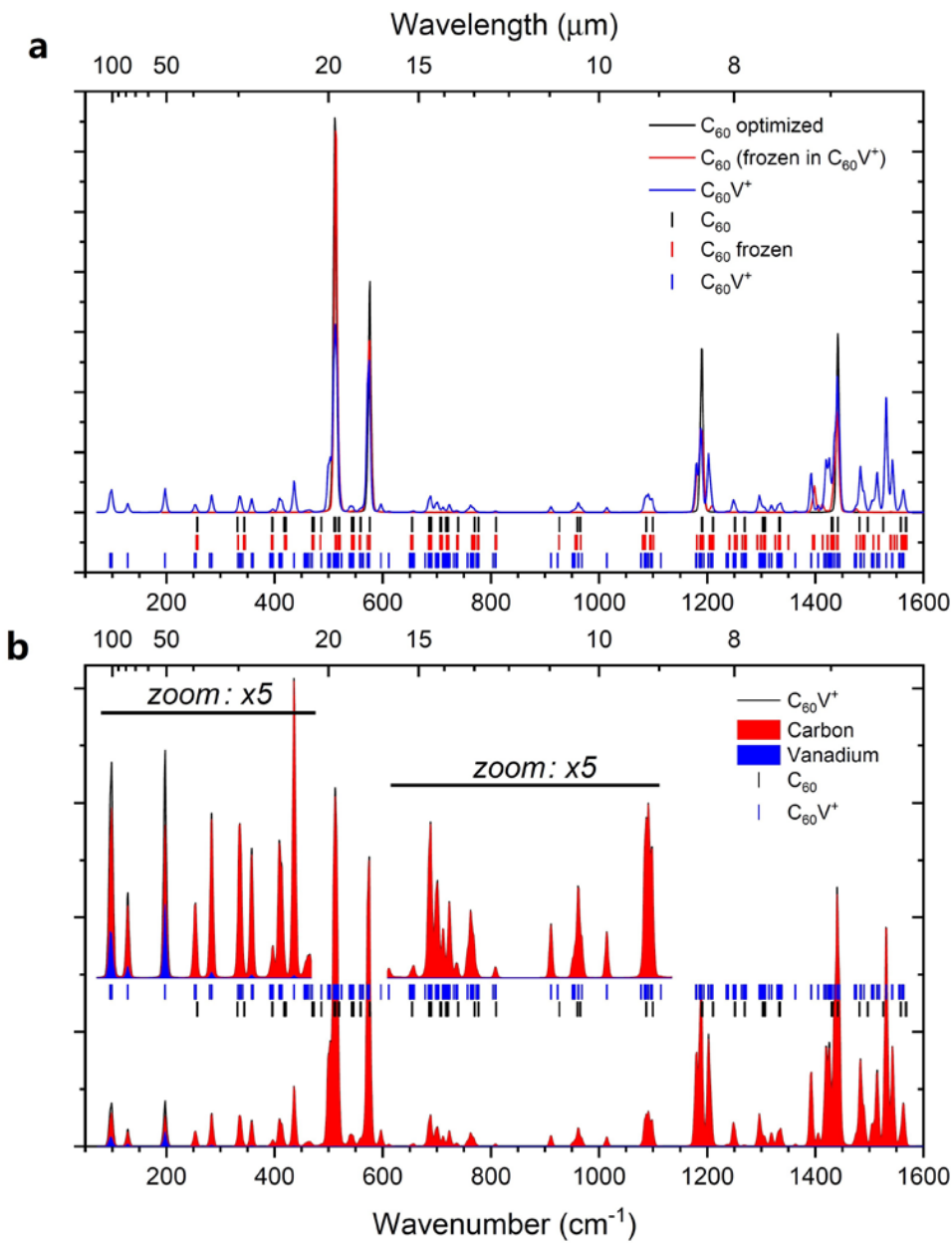

**Supplementary Figure S2. Influence of a metal atom on the vibrational property of C₆₀. a.** Vibrational spectra of the optimized icosahedral $C_{60}$ (black line), the "frozen" $C_{60}$ with the same geometry in $C_{60}V^+$ where it is distorted by the V atom to a lower symmetry (red line), and $C_{60}V^+$. The vibrational frequencies of all modes, including those carrying no infrared intensities are displayed in sticks using the same color convention. It can be clearly seen that the presence of metal atom introduces much more infrared intensity than just lowering the $C_{60}$ symmetry in addition to a few new vibrational modes. **b.** Spectral decomposition of vibrational modes into motions involving the V atom (blue) and the carbon cage (red). This data represents the relative vector magnitudes of the V and C components of each normal mode displacement vector, projected onto the convolved infrared spectrum for $C_{60}V^+$ (supplementary Figure S2a). It can be seen that vibrational modes above 500 cm⁻¹ (below 20 μm) are dominated by $C_{60}$ cage motions. Below 500 cm⁻¹ (above 20 μm), the V atom gets more involved, yielding metal-specific vibrational frequencies.

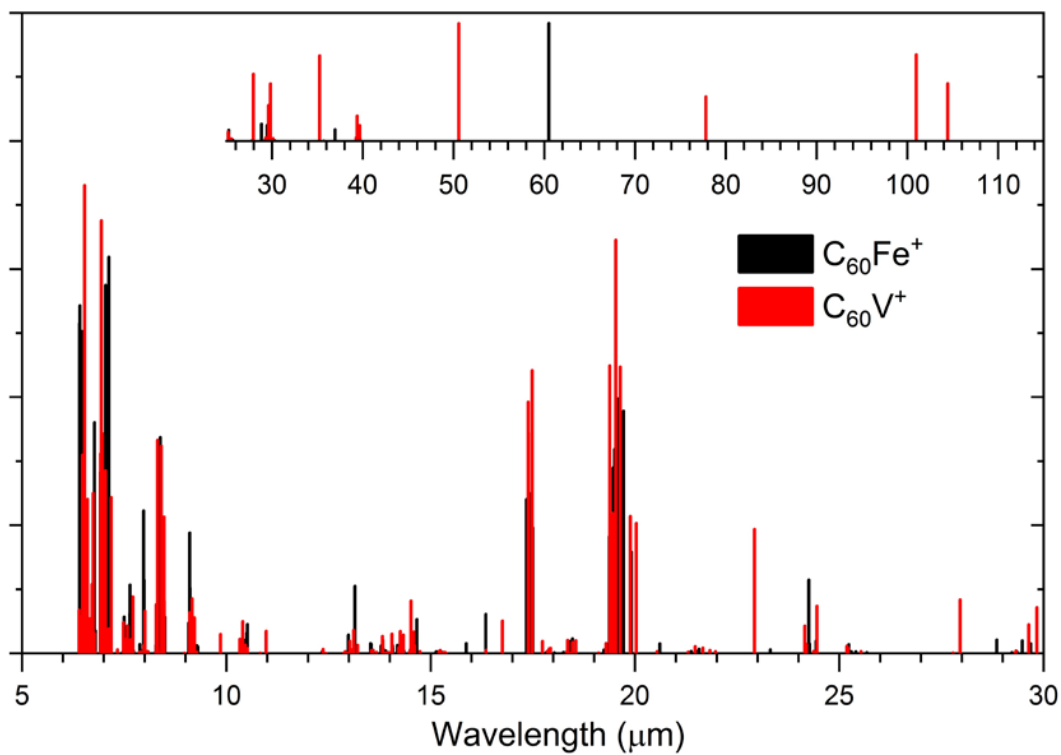

**Supplementary Figure S3 | Detailed comparison of the predicted infrared spectra of [C₆₀-Metal]⁺ (Metal = V, Fe).** It demonstrates that the metal-specific vibrational bands start to emerge beyond 20 μm, and become more metal-dependent beyond 30 μm as seen in supplementary Figure S1 as well.

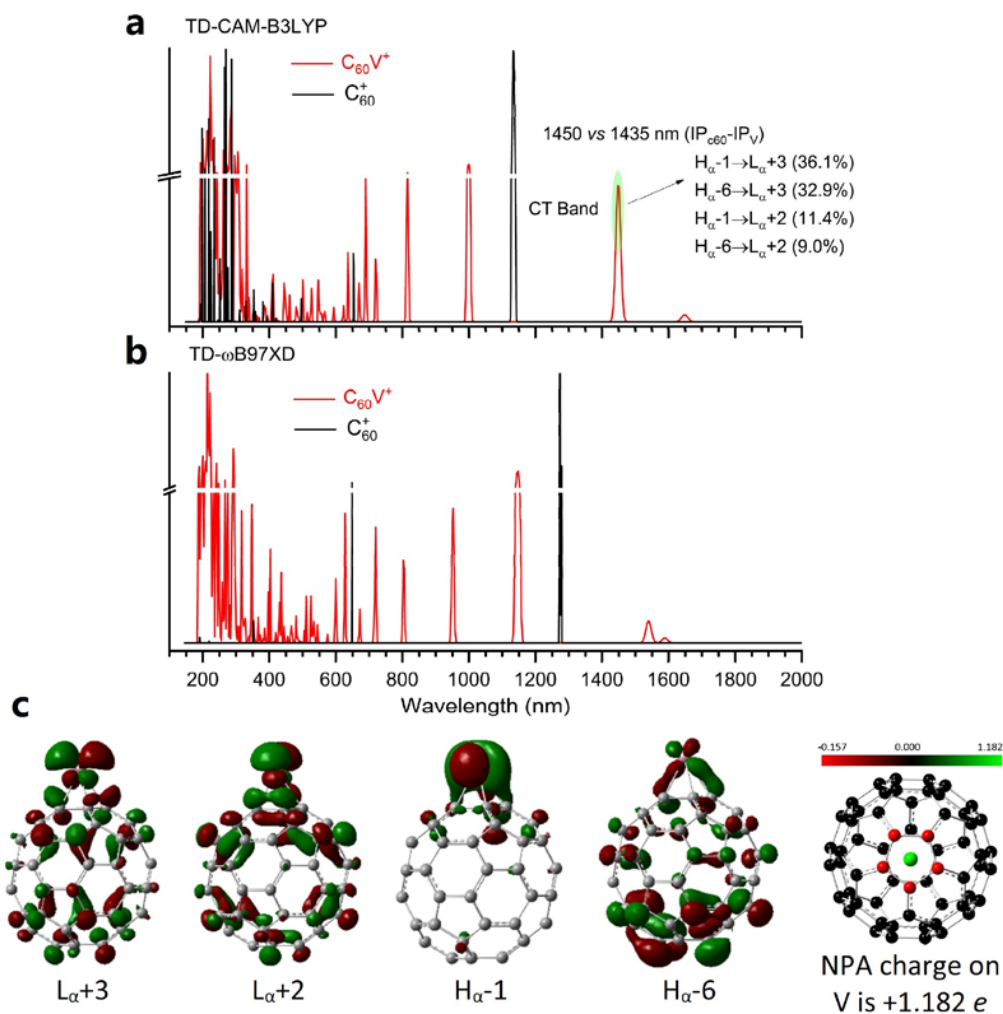

**Supplementary Figure S4 | Simulated ultraviolet to visible to near infrared (UV-Vis-NIR) absorption spectra of the most stable $C_{60}V^+$ using time-dependent density functional theory (TDDFT).** The molecular orbitals (MOs) involved in the transition at 1450 nm (charge transfer from metal to the $C_{60}$ cage band, consistent with the charge distributions from natural population analysis (NPA)) are plotted. $H_\alpha$ represents the highest occupied MO with spin up electron, and $L_\alpha$ represents the lowest unoccupied MO with spin up electron. Kroto et al. has anticipated that the [$C_{60}$-Metal]$^+$ complexes should exhibit strong charge transfer (CT) transitions with energies approximate to the ionization potential (IP) differences of $C_{60}$ and metals (Kroto et al. *Astron. Astrophys.* **263**, 275-280 (1992)). That will give a CT band for $C_{60}V^+$ to be at (7.61−6.746 = 0.864 eV) 1435 nm, close to the calculated CT band at 1450 nm (TD-CAM-B3LYP level of theory). The CT from the metal atom to the cage results in strong $C_{60}$-metal interaction. The CAM-B3LYP functional comprises of 0.19 Hartree–Fock (HF) plus 0.81 Becke 1988 (B88) exchange interaction at short-range, and 0.65 HF plus 0.35 B88 at long-range. The intermediate region is smoothly described through the standard error function with parameter 0.33 (Yanai et al. *Chem. Phys. Lett.* **393**, 51-57 (2004)). The CAM-B3LYP functional has been shown to be superior to describe the charge-transfer excitation transitions (Jacquemin et al. *J. Chem. Theory Comput.* **4**, 123-135 (2008)). The calculation using TD-ωB97XD is also carried out for comparison, and both functionals qualitatively show that $C_{60}V^+$ complex has much more electronic transitions in 400–1000 nm compared to $C_{60}^+$. Note that the quantitative prediction of the exact positions of those transitions is a huge challenge even for $C_{60}^+$ (Lykhin et al. *J. Phys. Chem. Lett.* **10**, 115-120 (2018); Soler et al. *J. Phys. Chem. A* **123**, 1824-1829 (2019)).

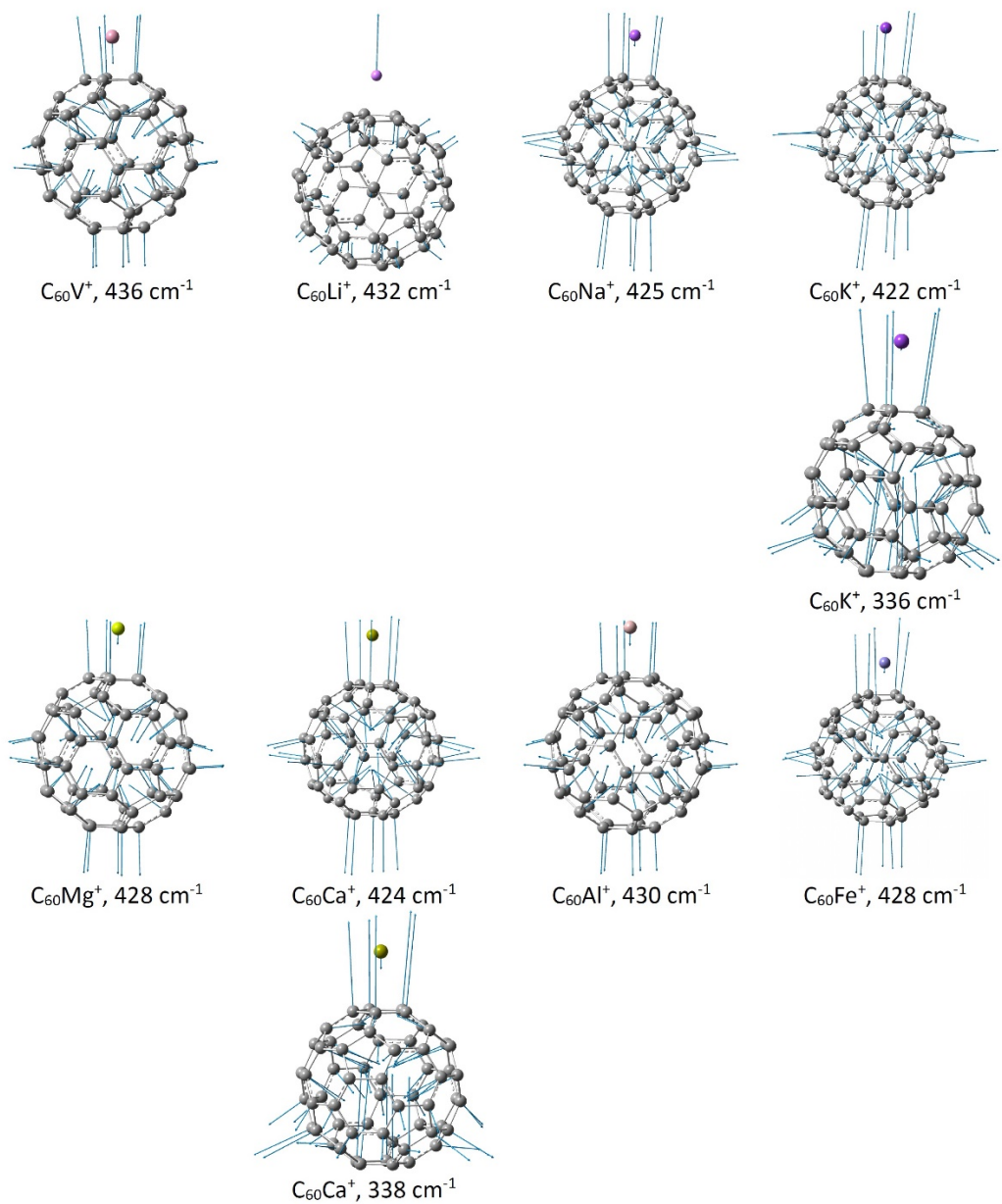

**Supplementary Figure S5 | The vectors of the metal–$C_{60}$ stretching modes of [$C_{60}$-Metal]$^+$ (Metal = V, Li, Na, K, Mg, Ca, Al, and Fe) and the calculated vibrational frequencies at BPW91/6-31G(d) level of theory.** Note that the $C_{60}K^+$ and $C_{60}Ca^+$ complexes have the metal-$C_{60}$ stretching modes at lower frequencies (~340 cm$^{-1}$); while their modes around 430 cm$^{-1}$ do not involve obvious movement of the metal atoms.

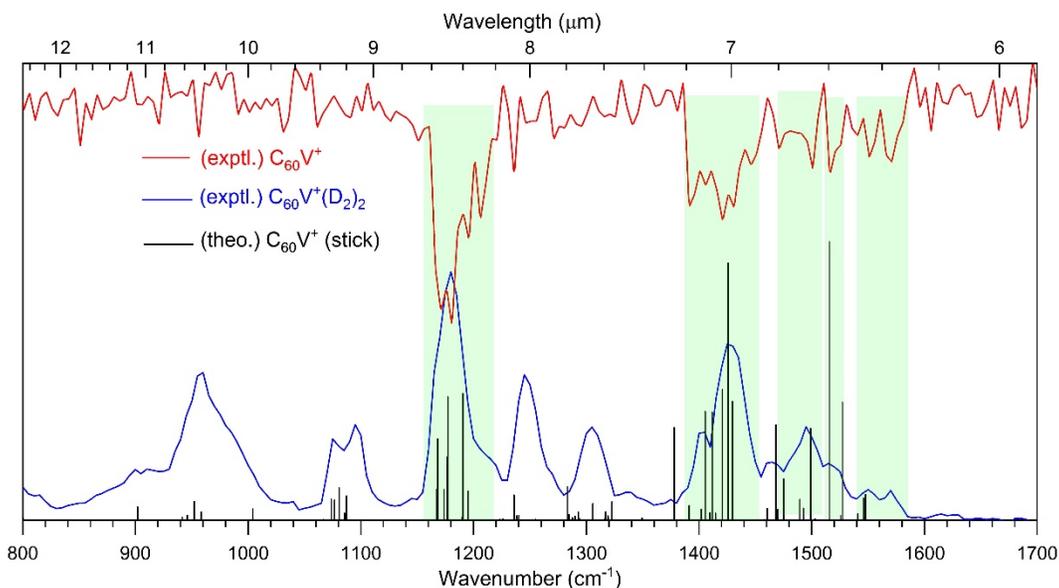

**Supplementary Figure S6 | Laboratory IRMPD spectrum of $C_{60}V^+$ and its comparison with that of two $D_2$-tagged $C_{60}V^+$ complexes and the DFT simulated stick spectrum of the most stable $C_{60}V^+$.** The IRMPD spectrum of $C_{60}V^+$ was obtained without $D_2$-tag. Hence, many photons are needed to induce the dissociation of $C_{60}V^+ \rightarrow C_{60} + V^+$ (the ionization potential of $C_{60}$ is 7.61 eV, and that of V is 6.746 eV; the dissociation energy is ca. 2.82 eV), and only the vibrational modes of high calculated infrared oscillator strengths were obtained. The features highlighted by green shading show that the $D_2$-tag almost does not alter the vibrational band positions of $C_{60}V^+$. The vibrational modes with low calculated oscillator strengths were not observed for $C_{60}V^+$, since at those wavelengths not enough infrared photons can be absorbed to reach the dissociation threshold.

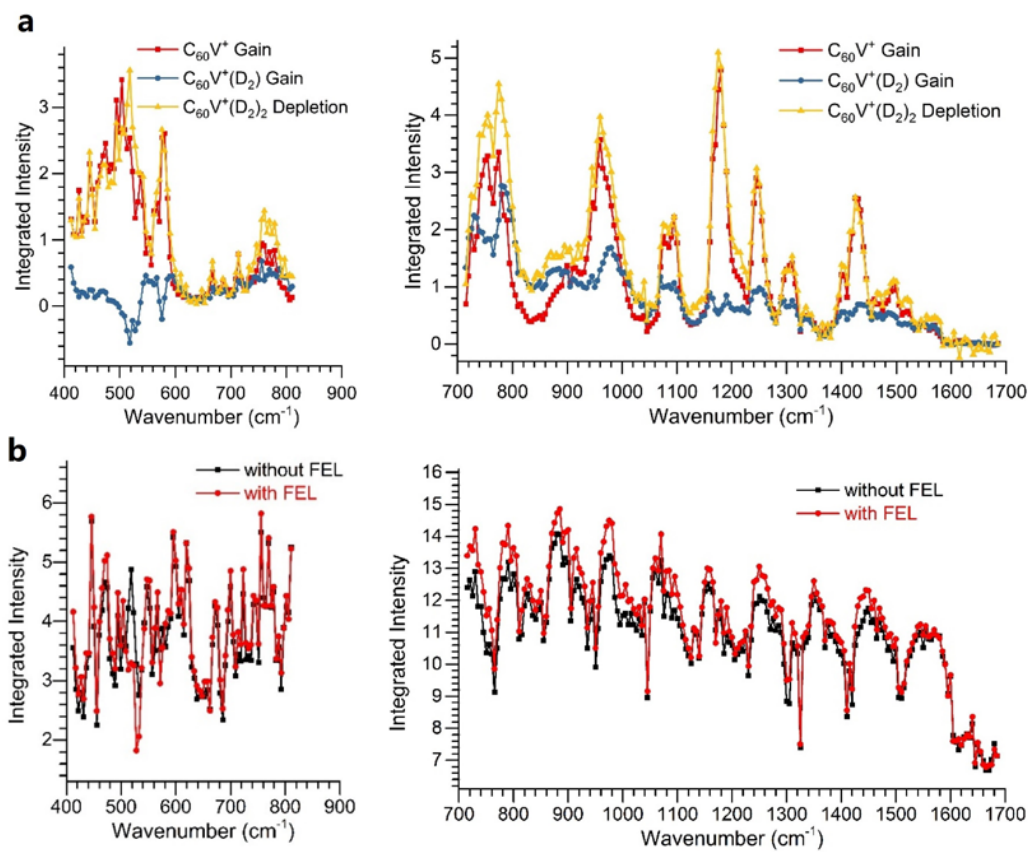

**Supplementary Figure S7 | Integrated ion intensity as a function of infrared wavelength in the range of 400–1700 cm⁻¹. a.** The changes of ion intensity of $C_{60}V^+$ and its $D_2$-tagged complexes. **b.** The total ion intensity of $C_{60}V^+$ and its $D_2$-tagged complexes with (red) and without (black) FEL light irradiation. In general, it can be seen that both $C_{60}V^+$ and $C_{60}V^+(D_2)$ gain intensities from the depletion of $C_{60}V^+(D_2)_2$, except around 520 and 570 cm⁻¹ where $C_{60}V^+$ also gains a little intensity from the depletion of $C_{60}V^+(D_2)$. The comparison of the summed ion intensities with and without FEL shows that the total ion intensity is conserved within the experimental uncertainty, implying that there is no other dissociation channel than $D_2$ loss. Note that the experiments were performed with two separate FEL settings, covering the 400–850 and 700–1700 cm⁻¹ ranges. The quasi-repetitive fluctuation in the signal is attributed to slight curvature in the target surface used for laser ablation, modulating the efficiency of the vaporization process.

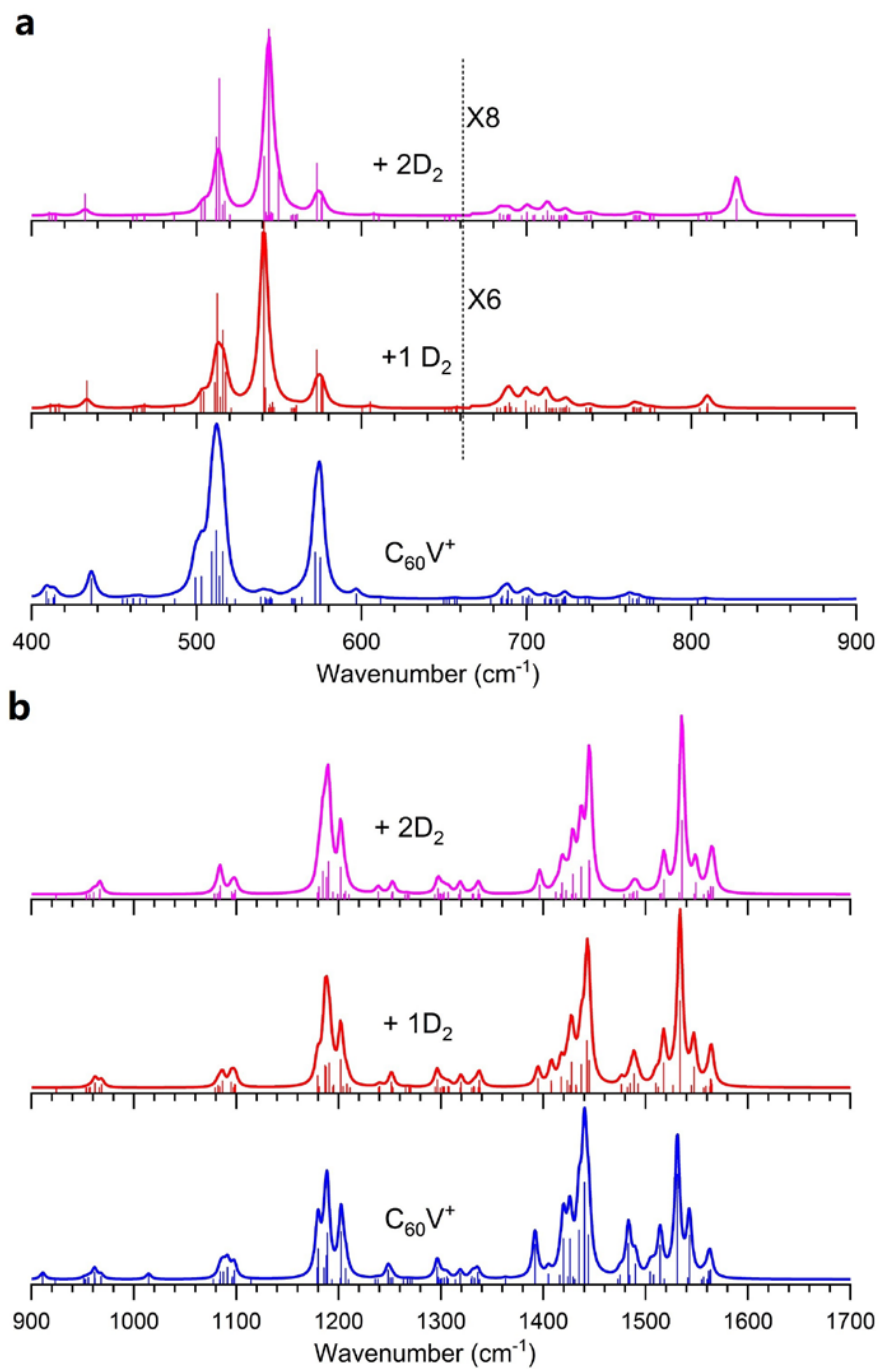

**Supplementary Figure S8 | Simulated harmonic infrared spectra of $C_{60}V^+(D_2)_{0-2}$ with stick lines showing each infrared mode.** The calculation is conducted at BPW91/6-31G(d) level of theory, and the convolved spectra are using Lorentzian line shapes of 6 cm⁻¹ full width at half maximum. The convolved spectra of $C_{60}V^+(D_2)$ and $C_{60}V^+(D_2)_2$ in the 600–900 cm⁻¹ region are multiplied by factors of 6 and 8, respectively. It can be seen that the main visible changes caused by $D_2$-tag are additional modes around 500–600 and 830 cm⁻¹. There are many closely spaced vibrational modes in the 500–600 cm⁻¹ range calculated for $C_{60}V^+(D_2)_2$, being one factor contributing to the broad band features observed in experiment. The vectors of the vibrational modes that significantly involve $D_2$ motions are provided in supplementary Figure S4.

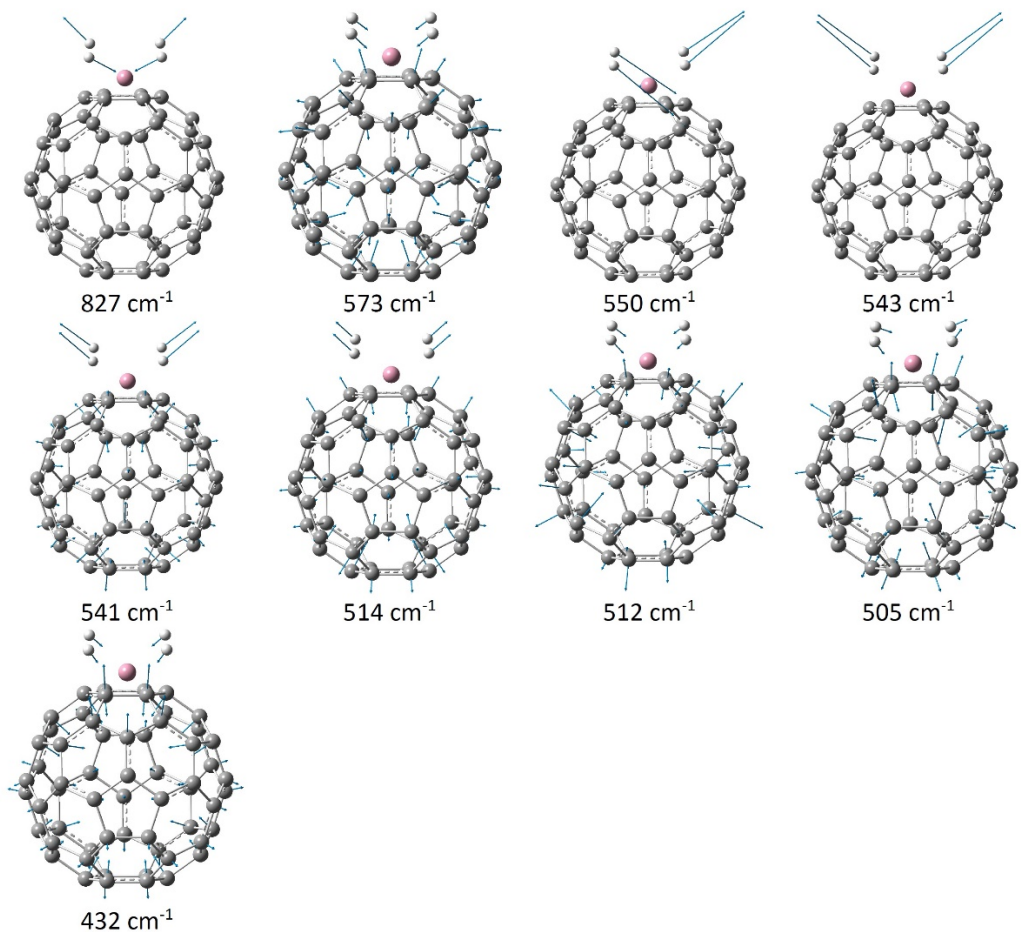

**Supplementary Figure S9 | Vectors of the vibrational modes significantly involving the motions of D₂ molecules** in the two D₂-tagged $C_{60}V^+$ complex, i.e., $C_{60}V^+(D_2)_2$, and the calculated vibrational frequencies at BPW91/6-31G(d) level of theory.

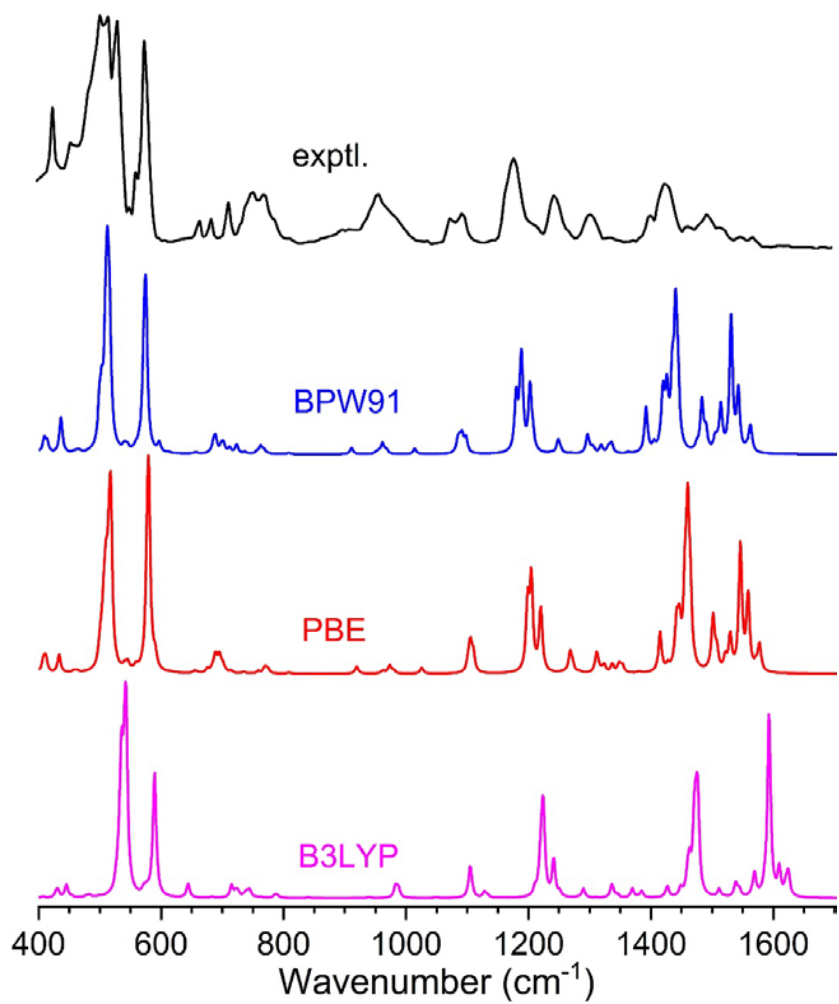

**Supplementary Figure S10 | Simulated infrared spectra of the most stable $C_{60}V^+$ using three different functionals and their comparison with the experimental IRMPD spectrum.** The simulated spectra are convolved using Lorentzian line shapes of 6 cm$^{-1}$ full width at half maximum. It can be seen overall the BPW91 functional provides the best agreement in comparison with experiment without applying any scaling factors.

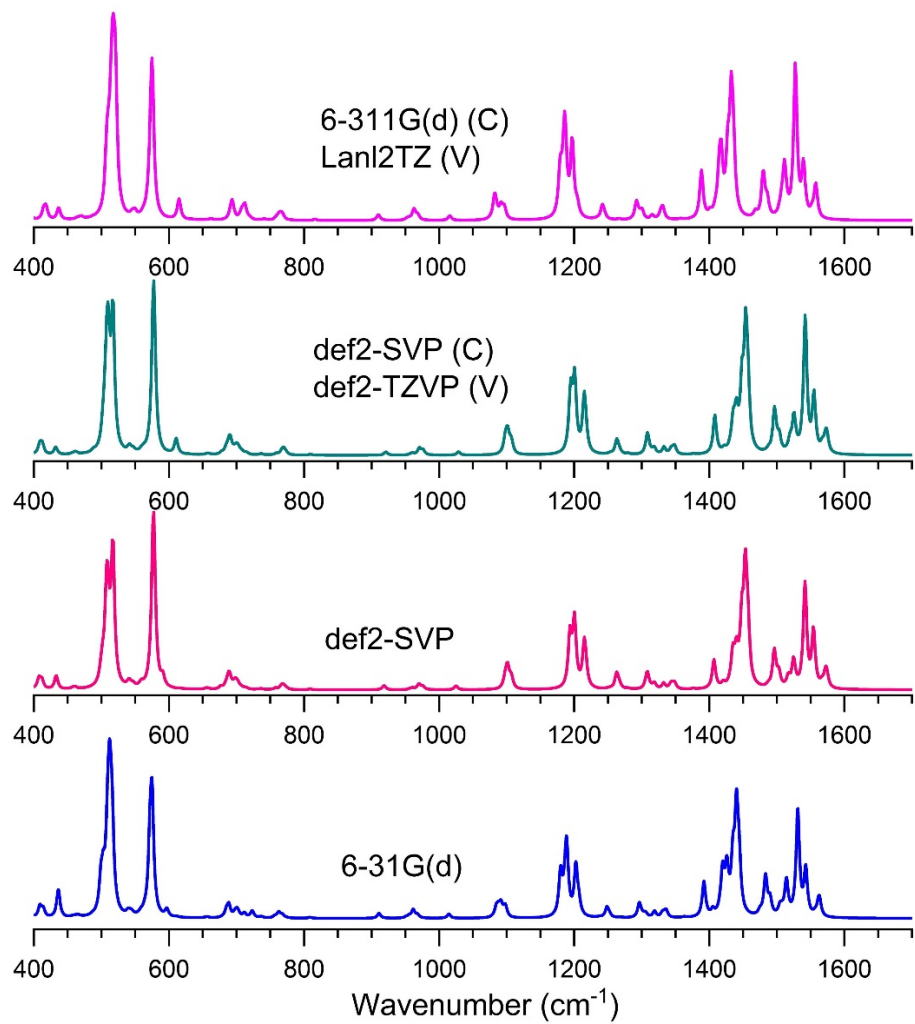

**Supplementary Figure S11 | Simulated infrared spectra of the most stable C₆₀V⁺ using the BPW91 functional with different basis sets.** The spectra are convolved using Lorentzian line shapes of 6 cm⁻¹ full width at half maximum. It can be seen that the basis sets have a minor effect on the simulated spectra regarding both the band positions and intensities.

**Supplementary Table S1.** The infrared spectral band positions from the laboratory experiment, the calculation of the most stable $C_{60}V^+$, and the Spitzer infrared spectra of Tc 1, LMC 56, SMC 6, and Lin49, as well as of $C_{60}$ and $C_{60}^+$. The asterisks indicate $D_2$-tag induced features. Note that the small band position differences between the laboratory measurement and the astronomical observations can well be accounted for by considering both the wavelength uncertainties in both the laboratory and Spitzer data. For instance, even for the four neutral $C_{60}$ bands, there is about 1% deviation of the band positions as seen in Tc 1.

| | $C_{60}V^+$ | | $C_{60}Fe^+$ | $C_{60}$ | Tc 1[a] | LMC 56[b] | SMC 16[b] | Lin49[c] | $C_{60}^+$[d] | $C_{60}$ | | $C_{70}$ | Atomic lines[a,b] |
|---|---|---|---|---|---|---|---|---|---|---|---|---|---|
| | Lab. | Theo. | Lab. | Theo. | | | | | | gas (~1065 K)[e] | solid (~300 K)[f] | Tc 1[a] | |
| | $cm^{-1}$/μm | $cm^{-1}$ | $cm^{-1}$/μm | $cm^{-1}$/μm | μm | μm | μm | μm | μm | $cm^{-1}$/μm | $cm^{-1}$/μm | μm | μm |
| 17-24 μm | | 412 | | | | | | 23.75 | | | | | |
| | 427/23.42 | 436 | 428/23.38 | | | | 23.31 | 23.33 | | | | | |
| | | | | | | | | | | | | | 22.92 [FeIII] |
| | 456/21.93 | | 436/22.90 | | 21.85 | 21.96 | | 21.88 | | | | 21.81 | 21.83 [ArIII] |
| | *504 | 500 | 486/20.58 | | | | | 19.67 | | | | | |
| | *518 | | 495/20.20 | | 19.30 | | | 19.41 | | | | | |
| | 532/18.80 | 512 | 531/18.83 | 512/19.53 | 18.90 | 18.83 | 18.82 | 18.95 | | 527.1/18.97 | 528/18.94 | 18.93 | |
| | | | | | | | | | | | | | 18.70 [SIII] |
| | | | | | | | 18.27 | 18.38 | | | | | |
| | 551/18.12 | 542 | 553/18.08 | | | 18.06 | 18.05 | 18.11 | | | | | |
| | 562/17.79 | 558 | 564/17.72 | | | 17.80 | 17.71 | 17.72 | | | | | 17.88 [PIII] |
| | 576/17.36 | 574 | 576/17.36 | 576/17.36 | 17.39 | 17.37 | 17.39 | 17.42 | | 570.3/17.53 | 577/17.33 | 17.38 | |
| | | | | | | 16.87 | 16.94 | | | | | | |
| | | | | | 16.22 | | 16.18 | 16.57 | | | | | |
| | | | | | | | 16.02 | | | | | | |
| | | | | | 15.55 | 15.57 | 15.58 | 15.35 | | | | 15.56 | 15.55 [NeIII] |
| 10-17 μm | 667/14.99 | 657 | 667/14.99 | | 14.78 | 14.80 | 15.00 | | | | | 14.79 | |
| | 686/14.58 | 688 | | | | 14.50 | 14.50 | | | | | | |
| | | 700 | 698/14.33 | | | | | 14.34 | | | | | 14.32 [NeV] |
| | 714/14.01 | 712 | | | | 14.05 | 14.07 | | | | | | |
| | | 724 | | | | | | 13.94 | | | | | |
| | | 737 | 737/13.56 | | | | | 13.73 | | | | | |
| | 754/13.26 | 763 | 757/13.20 | | | 13.29 | 13.25 | 13.23 | | | | | |
| | *773 | | | | 12.84 | 12.84 | 12.84 | 12.82 | | | | | 12.81 [NeII] |
| | 793/12.61 | | 787/12.70 | | 12.52 | 12.56 | 12.62 | | | | | 12.52 | |
| | 815/12.28 | 808 | 823/12.14 | | 12.35 | 12.35 | 12.33 | 12.35 | | | | | 12.37 [H_re] |
| | 897/11.16 | 911 | | | | 11.28 | 11.26 | 11.31 | | | | | 11.30 [H_re] |
| | | | | | | | 10.68 | 10.73 | | | | | |
| | 960/10.42 | 961 | 955/10.47 | | | 10.50 | 10.51 | 10.02 | 10.5 | | | | 10.50 [SIV] |
| 6-10 µm | 1075/9.30 | 1090 | 1074/9.31 | | | | 9.24 | | | | | | |
| | 1096/9.12 | 1098 | 1095/9.13 | | 9.00 | 9.01 | 9.03 | | | | | 8.88 (not detected) | 8.99 [ArIII] |
| | 1180/8.47 | 1180 | 1172/8.53 | 1191/8.40 | 8.50 | 8.50 | 8.49 | 8.46 | | 1169.1/8.55 | 1183/8.45 | | |
| | | 1189 | 1180/8.47 | | | | | | | | | | |
| | 1215/8.23 | 1202 | | | | | | | | | | | |
| | 1245/8.03 | 1248 | 1243/8.05 | | | 8.13 | 8.13 | 8.1 | | | | | |
| | | | 1269/7.89 | | | 7.94 | 7.95 | 8.08 | | | | | |
| | 1306/7.66 | 1297 | 1295/7.71 | | 7.68 | 7.66 | 7.64 | | | | | | |
| | | 1320 | | | | | | | | | | | |
| | 1340/7.46 | 1335 | 1354/7.38 | | 7.48 | 7.45 | 7.41/7.48 | 7.52 | | | | | 7.47 [H_re] |
| | | 1362 | | | | | | | | | | | |
| | 1402/7.13 | 1392 | | | | | 7.18 | 7.1 | | | | | |
| | | 1406 | | | | | | | | | | | |
| | | 1420 | | | | | | | | | | | |
| | 1426/7.01 | 1426 | 1427/7.00 | 1440/6.95 | 7.03 | 7.02 | 7.02 | 7.04 | | 1406.9/7.11 | 1429/7.00 | | 7.03 [ArII] |
| | 1465/6.83 | 1440 | 1467/6.82 | | | | 6.82 | | | | | | |
| | 1496/6.68 | 1485 | 1480/6.74 | | 6.67 | 6.65 | | | | | | | |
| | 1520/6.58 | 1514 | 1518/6.59 | | | | 6.58 | 6.59 | | | | | |
| | | 1531 | | | | | | | | | | | |
| | 1550/6.45 | 1542 | 1550/6.45 | | 6.48 | 6.47 | | 6.4 | | | | | 6.48 [H_re] |
| | 1570/6.37 | 1562 | 1572/6.35 | | 6.23 | 6.23 | | | | | | | |

**Supplementary Table S2.** The relative stabilities (in kJ/mol) and average V–C distances (in Å) of the different structures of $C_{60}V^+$ in different spin states at three different levels of theory, i.e., BPW91, PBE, and B3LYP. For a certain isomer and theory level combination, the optimization converged to a more stable isomer as indicated.

| | | BPW91/6-31G(d) | | PBE/def2-SVP | | B3LYP/def2-SVP | |
|---|---|---|---|---|---|---|---|
| | | ΔE | V–C | ΔE | V–C | ΔE | V–C |
| $\eta^5$ | S = 0 | 128.0 | 2.140 | 140.2 | 2.153 | 164.5 | 2.243 |
| 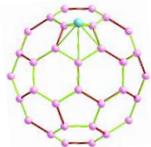 | S = 1 | 46.9 | 2.250 | 46.4 | 2.267 | 67.0 | 2.328 |
| | S = 2 | **0.0** | 2.282 | **0.0** | 2.306 | **0.0** | 2.416 |
| $\eta^6$ | S = 0 | 138.0 | 2.220 | 145.7 | 2.229 | 157.8 | 2.350 |
| 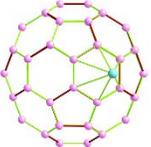 | S = 1 | 39.3 | 2.246 | 41.3 | 2.263 | 51.4 | 2.585 |
| | S = 2 | 7.5 | 2.358 | 11.5 | 2.419 | 0.9 | 2.510 |
| $\eta^{2(6\text{-}6)}$ | S = 0 | 207.8 | 1.912 | →$\eta^6$ | | 177.2 | 2.056 |
| 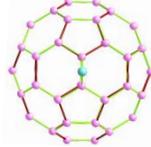 | S = 1 | →$\eta^6$ | | →$\eta^6$ | | 55.1 | 2.078 |
| | S = 2 | 31.4 | 2.181 | 33.5 | 2.202 | 3.59 | 2.290 |
| $\eta^{2(6\text{-}5)}$ | S = 0 | →$\eta^6$ | | →$\eta^6$ | | →$\eta^6$ | |
| 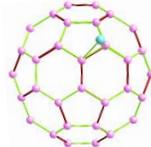 | S = 1 | 91.5 | 2.077 | →$\eta^5$ | | 54.5 | 2.068 |
| | S = 2 | 35.4 | 2.174 | 40.5 | 2.189 | →$\eta^6$ | |

**Supplementary Table S3.** Abundance of metal elements studied in this work. They are provided by the logarithm ratios of metals relative to the abundance of atomic hydrogen, i.e., $\lg(n_{metal}/n_H)$.

| Metals | Sol[a] | IS[a] | RedRec[b] |
|--------|--------|-------|-----------|
| H | 0 | 0 | 0 |
| C | -3.5 | -3.8 | -3.34 |
| V | -8.0 | -9.9 | — |
| Li | -8.7 | -10.3 | — |
| Na | -5.7 | -6.6 | — |
| K | -6.9 | -8.0 | — |
| Mg | -4.4 | -6.0 | -6.49 |
| Ca | -5.7 | -9.4 | -8.75 |
| Al | -5.6 | -7.9 | — |
| Fe | -4.5 | -6.8 | -7.82 |

[a]Solar and typical cool gas-phase interstellar abundances. From Jenkins, E. B. *Astrophys. J.* **700**, 1299-1348 (2009) and Savage, B. D. & Sembach, K. R. *Annu. Rev. Astron. Astrophys.* **34**, 279-329 (1996). [b]Photospheric abundances in the Red Rectangle. From Waelkens, C., Van Winckel, H., Trams, N. R. & Waters, L. B. F. M. *Astron. Astrophys.* **256**, L15-L18 (1992).

**Supplementary Table S4**. The abundances of fullerenes (percent of gas-phase carbon locked in fullerene species), derived from emission or absorption measurements in star forming regions, diffuse ISM, and evolved stars.

| | Star-forming regions | | Diffuse ISM | | Evolved stars | |
|---|---|---|---|---|---|---|
| | Emission | Absorption | Emission | Absorption | Emission | Absorption |
| $C_{60}^{+}$ | 0.01[a] | — | 0.2[c] (upper limit) | 0.06–0.1[c] | — | 1.2[e] |
| $C_{60}$ | 0.04–0.06[b] | — | 0.03–0.4[c] | — | 0.1–3.0[d] | — |

**Supplementary Table S5**. Collision Constant and thermal dissociation rate of $[C_{60}\text{-Metal}]^+$ complexes.

| Metals | Radius[a] (Å) | Polarizability[b] (Å³) | Binding Energy (eV) | Collision Rate Constant[c] ($10^{-9}$ cm³·s⁻¹) $k_{HS}$ 50 K | $k_{HS}$ 300 K | $k_{L1}$ | $k_{L2}$ | Formation Rate[d] (s⁻¹ per $C_{60}^+$) Sol | IS | Thermal Dissociation Rate[e] (s⁻¹) $k_d$ 50 K | $k_d$ 300 K |
|---|---|---|---|---|---|---|---|---|---|---|---|
| V | 1.71 | 12.9 | 2.82 | 0.22 | 0.54 | 3.03 | 1.22 | $2.45 \times 10^{-13}$ | $3.10 \times 10^{-15}$ | 0 | 0 |
| Li | 1.67 | 24.3 | 1.55 | 0.57 | 1.40 | 7.95 | 4.38 | $1.75 \times 10^{-13}$ | $4.40 \times 10^{-15}$ | 0 | $1.58 \times 10^{-11}$ |
| Na | 1.86 | 24.1 | 1.05 | 0.34 | 0.82 | 4.43 | 2.43 | $9.70 \times 10^{-11}$ | $1.23 \times 10^{-11}$ | 0 | $3.98 \times 10^{-3}$ |
| K | 2.43 | 42.9 | 0.67 | 0.31 | 0.75 | 3.44 | 2.52 | $6.35 \times 10^{-12}$ | $5.05 \times 10^{-13}$ | 0 | $7.94 \times 10^{3}$ |
| Mg | 1.45 | 10.6 | 1.07 | 0.29 | 0.71 | 4.34 | 1.58 | $1.25 \times 10^{-9}$ | $3.16 \times 10^{-11}$ | 0 | $2.00 \times 10^{-3}$ |
| Ca | 1.94 | 23.8 | 1.07 | 0.26 | 0.65 | 3.40 | 1.85 | $7.40 \times 10^{-11}$ | $1.48 \times 10^{-14}$ | 0 | $2.00 \times 10^{-3}$ |
| Al | 1.18 | 8.6 | 1.69 | 0.25 | 0.62 | 4.10 | 1.35 | $6.76 \times 10^{-11}$ | $3.40 \times 10^{-13}$ | 0 | $6.31 \times 10^{-14}$ |
| Fe | 1.56 | 9.2 | 2.25 | 0.20 | 0.49 | 2.90 | 0.98 | $6.23 \times 10^{-10}$ | $3.10 \times 10^{-12}$ | 0 | 0 |

[a]The atomic radii are from: https://en.wikipedia.org/wiki/Atomic_radii_of_the_elements_(data_page).
[b]The polarizabilities are from Schwerdtfeger, P. & Nagle, J. K. *Mol. Phys.* **117**, 1200-1225 (2019).
[c]$k_{HS}$ is the collision rate constant for the neutral route $C_{60}$ + Metal, and $k_{L1}$ and $k_{L2}$ are for $C_{60}$ + Metal⁺ and $C_{60}^+$ + Metal routes, respectively.
[d]The formation rates of $[C_{60}\text{-Metal}]^+$ complexes via $C_{60}^+$ + Metal route for a single $C_{60}^+$ molecule based on the solar and interstellar metal abundances, respectively (supplementary Table S3).
[e]The universe has an estimated age of 13.8 billion years (Planck Collaboration, *A&A* **594**, A13 (2016)), therefore we define the dissociate rates lower than the age of Universe, i.e., $1.38 \times 10^{10}$ yrs or $4.35 \times 10^{17}$ s, to be zero.